\documentclass[onecolumn,fleqn,usenatbib]{mnras}

\usepackage{newtxtext,newtxmath}

\usepackage[T1]{fontenc}
\DeclareRobustCommand{\VAN}[3]{#2}
\let\VANthebibliography\thebibliography
\def\thebibliography{\DeclareRobustCommand{\VAN}[3]{##3}\VANthebibliography}

\usepackage{float}
\usepackage{ulem}
\usepackage{relsize}
\pdfoutput=1
\usepackage{microtype}
\usepackage{natbib,caption}             
\usepackage{threeparttablex}
\usepackage{tabularx}                   
\usepackage{booktabs,amsfonts}          
\usepackage{graphicx}	                  
\usepackage{amsmath}	               
\usepackage{color}

\title[Binary UCD Radiation Belt Occurrence Rate]{Binarity Enhances the Occurrence Rate of Radiation Belt Emissions in Ultracool Dwarfs}

\author[Kao \& Pineda]{
Melodie M. Kao,$^{1,2,3,4,5}$\thanks{E-mail: mkao@lowell.edu}
J. Sebastian Pineda$^{6}$
\\
$^{1}$Arizona State University, School of Earth and Space Exploration, 550 E Tyler Mall PSF 686, Tempe, AZ 85287\\
$^{2}$University of California Santa Cruz, Department of Astronomy amd Astrophysics, 1156 High Street, Santa Cruz, CA 95064\\
$^{3}$Lowell Observatory, 1400 W Mars Hill Road, Flagstaff, AZ 86001\\
$^{4}$NASA Hubble Postdoctoral Fellow\\
$^{5}$Heising-Simons 51 Pegasi b Fellow\\
$^{6}$University of Colorado Boulder, Laboratory for Atmospheric and Space Physics, 3665 Discovery Drive, Boulder CO, 80303, USA
}

\date{Accepted 2024 March 12. Received YYY; in original form ZZZ}

\pubyear{2024}

\begin{document}
\label{firstpage}
\pagerange{\pageref{firstpage}--\pageref{lastpage}}
\maketitle

\begin{abstract}
Despite a burgeoning set of ultracool dwarf ($\leq$M7) radio detections, their radio emissions remain enigmatic.  Open questions include the plasma source and acceleration mechanisms for the non-auroral ``quiescent" component of these objects' radio emissions, which can trace Jovian synchrotron radiation belt analogs. Ultracool dwarf binary systems can provide test beds for examining the underlying physics for these plasma processes.  We extend a recently developed occurrence rate calculation framework to compare the quiescent radio occurrence rate of binary systems to single objects. This generalized and semi-analytical framework can be applied to any set of astrophysical objects conceptualized as unresolved binary systems with approximately steady-state emission or absorption. We combine data available in the literature to create samples of 179 single ultracool dwarfs (82 M dwarfs, 74 L dwarfs, and 23 T/Y dwarfs) and 27 binary ultracool dwarf systems.  Using these samples, we show that quiescent radio emissions occur in $54^{+11}_{-11}$ -- $65^{+11}_{-12}$ per cent of binaries where both components are ultracool dwarfs, depending on priors.  We also show that binarity enhances the ultracool dwarf quiescent radio occurrence rate relative to their single counterparts.    Finally, we discuss potential implications for the underlying drivers of  ultracool dwarf quiescent radio emissions, including possible plasma sources.

\end{abstract}




\section{Introduction}\label{sec:intro}

Magnetic activity undergoes a fundamental shift as ultracool dwarfs (M7 and later spectral types) drop in effective temperature and become more analogous to gas giant planets. Ultracool dwarfs at late as $\sim$T8 \citep{RouteWolszczan2012ApJ...747L..22R, Kao2016ApJ...818...24K, Williams2017ApJ...834..117W, Vedantham2020ApJ...903L..33V, Vedantham2023A&A...675L...6V, Rose2023ApJ...951L..43R} can produce bursting radio emissions that are orders of magnitude stronger than predicted by coronal X-ray emissions \citep[e.g.][]{Berger2001Natur.410..338B, Williams2014ApJ...785....9W, Pineda2017ApJ...846...75P, Callingham2021NatAs...5.1233C}. The trio of papers establishing the auroral paradigm of magnetic activity in the ultracool dwarf regime \citep{Hallinan2015Natur.523..568H, Kao2016ApJ...818...24K, Pineda2017ApJ...846...75P} argued that their coherent and periodically bursting radio emissions, emitted via the electron cyclotron maser instability \citep{Hallinan2007ApJ...663L..25H, Hallinan2008ApJ...684..644H}, trace the radio component of aurorae in these objects. Searches for such coherent emissions from ultracool dwarfs have enabled the only direct measurements of magnetic field strengths on planetary-mass objects outside of the Solar System \citep{Kao2016ApJ...818...24K, Kao2018ApJS..237...25K}, yielded the first direct discovery of a brown dwarf using radio observations \citep{Vedantham2020ApJ...903L..33V}, confirmed that late T dwarfs can host kiloGauss magnetic fields \citep{RouteWolszczan2012ApJ...747L..22R, RouteWolszczan2016ApJ...821L..21R, Williams2015ApJ...808..189W, Kao2016ApJ...818...24K, Kao2018ApJS..237...25K, Rose2023ApJ...951L..43R}, and demonstrated that our understanding of dynamos in the mass and temperature regime bridging low mass stars and planets remains limited \citep{Kao2016ApJ...818...24K, Kao2018ApJS..237...25K}.

Ultracool dwarfs can also produce long-lived incoherent and quasi-steady nonthermal radio emissions \citep[e.g.][and references therein]{Hallinan2007ApJ...663L..25H, Kao2016ApJ...818...24K, Kao2023Natur.619..272K,  Guirado2018AA...610A..23G, climent2022AA...660A..65C, Climent2023Sci...381.1120C, kao2024MNRAS.527.6835K}, which extends to at least 95~GHz \citep{Williams2015ApJ...815...64W, hughes2021AJ....162...43H} and which numerous studies attribute to optically thin gyrosynchrotron emissions \citep[e.g.][]{Berger2005ApJ...627..960B, Osten2006ApJ...637..518O, Williams2015ApJ...815...64W, Lynch2016MNRAS.457.1224L}. \citet{Williams2015ApJ...815...64W} proposed that a blanket of low-level flaring could be one possible explanation, and indeed strong white-light flares can occur on objects as late as early to mid-L dwarfs \citep[e.g.][]{Gizis2013ApJ...779..172G, Paudel2018ApJ...858...55P, Jackman2019MNRAS.485L.136J, Paudel2020MNRAS.494.5751P, Medina2020ApJ...905..107M}.  At present, the possibility of flare behavior extending to cooler objects remains unknown, as is also the case regarding the existence and rate of weaker radio flares in the coldest T- and Y-spectral type brown dwarfs.  

However, existing studies show that flare activity may not correlate with radio activity on ultracool dwarfs.  Flares become less frequent with later spectral types \citep{Paudel2018ApJ...858...55P, Medina2020ApJ...905..107M}, and emission strengths from the chromospheric activity marker H$\alpha$ also decrease  \citep{Schmidt2015AJ....149..158S, Pineda2017ApJ...846...75P}.  If flares indeed correlate with quiescent radio emissions, we expect the latest type ultracool dwarfs to similarly exhibit a lower radio occurrence rate when all other factors are equal. Instead, T/Y dwarfs do not exhibit a lower quiescent radio occurrence rate than M dwarfs  \citep{kao2024MNRAS.527.6835K}. Though these results do not control for existing biases in science that prioritize publishing detections over non-detections, nor for other possible confounding factors such as age or rotation rate, they are consistent with other studies finding that ultracool dwarf quiescent radio emissions misalign with flare behaviors. For instance, quiescent radio luminosities from ultracool dwarfs depart \citep{Williams2014ApJ...785....9W, Pineda2017ApJ...846...75P} from well-established X-ray vs. radio correlations for stellar flare activity \citep{Guedel1993ApJ...405L..63G}. Instead, for auroral emitters, they correlate with H$\alpha$ luminosities \citep{Pineda2017ApJ...846...75P, Richey-Yowell2020} attributed to aurorae \citep{Hallinan2015Natur.523..568H, Kao2016ApJ...818...24K, Pineda2017ApJ...846...75P}. Together, these findings suggest that flare activity cannot account for quiescent ultracool dwarf radio emissions. 

\citet{Hallinan2006ApJ...653..690H} postulated an alternative interpretation in which these objects' quiescent radio emissions trace dense regions of energetic particles trapped in their large-scale magnetospheres, similar to the radiation belts possessed by magnetized Solar System planets \citep{sault1997, bolton2004, Clarke2004jpsm.book..639C, Horne2008, MaukFox2010JGRA..11512220M}. In later work, \citet{Pineda2017ApJ...846...75P} and \citet{Kao2019MNRAS.487.1994K} developed this idea into a coherent framework tying together the auroral and quiescent components of ultracool dwarf radio emissions. Using multi-epoch imaging at 8.4~GHz, \citet{Kao2023Natur.619..272K} demonstrated that quiescent radio emissions from the canonical auroral ultracool dwarf LSR~J1835+3259 traces synchrotron instead of gyrosynchrotron emissions from $\sim$15~MeV electrons in Jovian radiation belt (or Earth van Allen belt) analogs. Notably, they showed that the observed radiation belts cannot trace electrons accelerated solely by flare processes. A later epoch of resolved imaging at 4.5~GHz confirm their result \citep{Climent2023Sci...381.1120C}. Together, these images show persistent equatorial belts of high energy plasma captured in the dipole magnetic field of LSR~J1835+3259 and extending further than nine times the object's radius despite the absence of stellar wind from a host star to seed its magnetospheric plasma.  All confirmed auroral ultracool dwarfs exhibit similar quiescent radio emissions \citep{Pineda2017ApJ...846...75P, Kao2016ApJ...818...24K, Richey-Yowell2020}, though whether such emissions can occur around objects without aurorae remains an open question. 

Recently, \citet{kao2024MNRAS.527.6835K} argued for quantifying ultracool dwarf magnetic activity using their quiescent emissions in addition to their periodically bursting aurorae.  They noted that the rarity of detected ultracool dwarf radio emissions --- between  $\sim$5--10 per cent detection rates for M, L, and T ultracool dwarfs \citep{Antonova2013AA...549A.131A, Lynch2016MNRAS.457.1224L, RouteWolszczan2016ApJ...830...85R} in volume-limited surveys --- together with the wealth of data available in the literature underscored the value of moving toward studies of radio occurrence rates as a function of object characteristics, such as $T_{\mathrm{eff}}$, mass,  rotation rate, and age. Occurrence rate studies also lend themselves well to all-sky surveys from current and upcoming instruments such as the LOw Frequency Array \citep[LOFAR\footnote{https://lofar-surveys.org};][]{vanHaarlem2013AA...556A...2V}, the Square Kilometre Array \citep[SKA\footnote{https://www.skao.int};][]{Braun2019arXiv191212699B}, and Deep Synoptic Array 2000 \citep[DSA-2000\footnote{https://www.deepsynoptic.org};][]{hallinan2019dsa2000}. Accordingly, \citet{kao2024MNRAS.527.6835K} developed a generalized semi-analytical framework for calculating occurrence rates for approximately steady-state emission or absorption in populations of single objects.  By applying this framework to observations available in the literature, they show that quiescent radio occurrence rates are between $15^{+4}_{-4}$ --  $20^{+6}_{-5}$ per cent for isolated ultracool dwarfs and that T/Y and M dwarfs exhibit similar radio occurrence rates. 

To date, no comprehensive study of ultracool dwarf radio emissions has focused on binaries, yet comparing the population of single objects to ultracool dwarf binaries provides a complementary means of identifying characteristics that influence magnetic activity at the stellar-planetary boundary.  For instance,  in stars, tidal spin-up may enhance the rotation rates and thus the magnetic activity of the binary population \citep{Zahn1977A&A....57..383Z, Morgan2016AJ....151..114M}. Here, we assess whether binarity enhances the quiescent radio occurrence rate of ultracool dwarfs in binary systems compared to their single counterparts and discuss implications.

\newcommand{\rms}{\sigma_{\mathrm{rms}}}
\newcommand{\rmsi}{\sigma_{\mathrm{rms_i}}}
\newcommand{\derr}{\sigma_{\mathrm{d}}}
\newcommand{\derri}{\sigma_{\mathrm{d_i}}}
\newcommand{\pierri}{\sigma_{\mathrm{\pi_i}}}
\newcommand{\pierr}{\sigma_{\mathrm{\pi}}}
\newcommand{\detect}{\mathrm{detect}}
\newcommand{\thr}{\mathrm{thr}}
\newcommand{\prob}{\mathbb{P}}
\newcommand{\PDF}{\mathcal{P}}
\newcommand{\norm}{\mathcal{N}}
\newcommand{\vin}{V_{\mathrm{in}}}
\newcommand{\vout}{V_{\mathrm{out}}}
\newcommand{\vtot}{V_{\mathrm{tot}}}
\newcommand{\bi}{\text{binary}}
\newcommand{\si}{\text{single}}
\newcommand{\predict}{\text{predict}}
\newcommand{\pop}{\mathcal{B}}


\section{Extending an occurrence rate framework for single objects to binary systems}\label{sec.Occurrence_Rate}

Does binarity affect the magnetic activity of ultracool dwarfs? To answer this question, we compare quiescent radio occurrence rates for ultracool dwarfs in single-object versus binary systems by adapting the occurrence rate framework developed by \citet{kao2024MNRAS.527.6835K}. We use their notation here, where $\prob$ denotes probability and $\PDF$ refers to a probability density distribution function. 

This Bayesian occurrence rate framework for a sample of single objects considers the probability $\theta_{\si}$ that an object is emitting between an assumed luminosity range $L \in [L_{\text{min}}, L_{\text{max}}]$ for
 a dataset $D$ consisting of individual observations where the $i^\mathrm{th}$ observation $D_i = \{ \detect_i, \rmsi, d_i, \derri, L_{i} \}$.  They treat each observation property as a random variable, where detect$_i \in \{0 \text{ if undetected}, 1 \text{ if detected}\}$, $\rmsi$ is the reported rms noise for each object, $d_i$ is the distance to each object, $\derri$ is the error for the measured $d_i$, and $L_{i}$ is the assumed specific luminosity for quiescent radio emission.  They show that for single-object systems, the probability $\prob(D \mid \theta_{\si})$ of observing a dataset given a fixed occurrence rate $\theta_{\si} = \Theta$ depends on the probability of drawing the particular observational and object properties captured in $D$:

\begin{equation} \label{eqn:p_Di_final}
\prob(D \mid \theta_{\si}=\Theta)= \displaystyle{\prod_{i=1}^N  }  \prob(\rmsi) \, \prob(\derri) \, \prob(d_i)  \times
\begin{cases} 
      \Theta \, \prob(L_{i} \geq L_{\mathrm{thr, i}} \mid \rmsi,\derri, d_i, e=1) 	 	&, \,\,   \detect_i = 1 \\
      1-\Theta \,  \prob(L_{i} \geq L_{\mathrm{thr, i}} \mid \rmsi,\derri, d_i, e=1)      &, \,\, \detect_i = 0 \quad, 
\end{cases}
\end{equation}
where  $\prob(L_{i} \geq L_{\mathrm{thr, i}} \mid \rmsi,\derri, d_i, e=1)$ is the probability  that the object has a detectable luminosity $L_{\mathrm{thr, i}}$, and  $e \in \{0=\text{not emitting},1=\text{emitting}\}$ denotes whether or not an object is emitting within the luminosity range of interest. Here, we follow \citet{kao2024MNRAS.527.6835K} and use a slight abuse of notation, where $\prob(\rmsi)$, $\prob(\derri)$, and $\prob(d_i)$ refer to the probabilities that an object possesses the given observations/properties within a representative physical interval of interest. We also use distance and parallax interchangeably, and \citet{kao2024MNRAS.527.6835K} provide the algebra for using either.  In the analysis that we present in \S \ref{sec:results_occurrenceRate}, we use parallaxes when available and distance estimates otherwise. 

To adapt the single-object framework to binary systems, we treat each system as an unresolved binary, as is the case for most observations reported in the literature. This approach allows us to account for the fact that we often do not know which component in the binary is emitting for detected systems. Thus, $\theta_{\si}$ becomes the occurrence rate for quiescent radio emissions in binary systems $\theta_{\bi}$. 

\citet{kao2024MNRAS.527.6835K} describe how to calculate $\prob(\rmsi)$, $\prob(\derri)$, and $\prob(d_i)$, which remain unchanged for unresolved binary systems. They additionally describe how to calculate $\prob(L_{i} \geq L_{\mathrm{thr, i}} \mid \rmsi,\derri, d_i, e=1)$ for single objects. To calculate this probability that an emitting object is also detectable, they integrate over the set of possible object distances. They also integrate over detectable object luminosities from an assumed luminosity distribution for emitting objects $\PDF(L_{\si} \mid e=1)$ within a luminosity range of interest  $L_{\si} \in [L_{\text{min}}, L_{\text{max}}]$: 
\begin{equation} 
		\prob(L_{i}  \geq  L_{\mathrm{thr,i}}   \mid  \rmsi,\derri, d_i, e=1) =  
		\displaystyle{\int\limits_{0}^{d_{\mathrm{thr},{\mathrm{max_i}}}}}   \PDF(x_i \mid \derri, d_i)  
		\displaystyle{\int\limits_{L_{\mathrm{thr, i}}}^{ L_{\max} } }   \PDF(L_{i} \mid e=1)  \quad dL_{i}  \,dx_i  \quad.
\end{equation}
 The only change required for applications to unresolved binary systems is replacing $\PDF(L_{\si} \mid e=1)$ with the luminosity distribution of unresolved binary systems given that the system is emitting, $\PDF(L_{\bi} \mid e=1)$, between an assumed luminosity range of interest $L_{\bi} \in [L_{\text{\bi, min}}, L_{\text{\bi, max}}]$.  Below, we define these minimum and maximum luminosities for binary systems. 
 
 For the remainder of this section, we address  $\PDF(L_{\bi} \mid e=1)$. To compare binary and single-object samples, we construct $\PDF(L_{\bi} \mid e=1)$ within the context of the luminosity distribution for single objects  $\PDF(L_{\si})$.  This is because unresolved binaries emit when one or both of their components are emitting.  Furthermore, they can satisfy the emission threshold even if both components are individually too faint to meet it (\S\ref{sec:changevar}).  In this framework,  we account for all cases.

The luminosity for an unresolved binary is equal to the summed luminosities of the $a$ and $b$ components such that $L_{\bi} = L_{a} + L_{b}$.  We assume that the luminosities for the individual components each follow the distribution $\PDF(L_{\si})$ , which gives the convolution
\begin{equation}\label{eqn:binary_luminosity}
	\PDF(  L_{\bi}) 	=	
		\begin{cases}
			\PDF(L_{\si}) * \PDF(L_{\si})  \, &, \,\,  0 \leq L_{\bi} \leq 2 L_{\mathrm{max}} \\
			0  \, &, \,\, \text{otherwise} \quad.
		\end{cases}
\end{equation}
Notice that this extends possible  luminosities for $\PDF(L_{\bi})$ to higher luminosities.  Accordingly, the binary luminosity range of interest for $e=1$ becomes $L_{\bi} \in [L_{\text{min}}, 2L_{\text{max}}]$, where we remind the reader that $L_{\text{max}}$ is the maximum luminosity for a single object. 

Here, the probability mass for  $\PDF(L_{\si})$ is distributed such that the occurrence rate of quiescent radio emissions for a single object, $\theta_{\si}$, is equal to the integrated probability mass between the minimum and maximum possible luminosities for a single object:
\begin{equation} \label{eqn:probability_mass}
    \int_{L_{\min}}^{L_{\max}} \PDF(L_{\si}) dL_{\si}= \theta_{\si}  \quad
\end{equation}
The remaining probability mass  $1-\theta_{\si}$  is distributed over  $L_{\si} \not\in [L_{\text{min}}, L_{\text{max}}]$. Notice that the combined luminosity of an unresolved binary depends on how probability mass is distributed both inside and outside the luminosity range of interest for single objects (see \S\ref{sec:lum_prior}).

When we consider the case that both components are emitting, the distribution of probability mass $1-\theta_{\si}$  outside of  $L_{\si} \not\in [L_{\text{min}}, L_{\text{max}}]$  becomes relevant.  For instance, an individual object that is emitting at $L_{\si} = 0.8 L_{\text{min}}$ is not considered ``emitting'' in the single-object framework.  In contrast, if both components in a binary are emitting at that luminosity, the binary luminosity  $L_{\bi}=  1.6L_{\text{min}}$ is within the binary luminosity range of interest.  In \S \ref{sec:lumDist}, we discuss our choice for distributing the non-emitting probability mass $1-\theta_{\si}$.

We can now re-write equation \ref{eqn:p_Di_final} to: 
\begin{equation} \label{eqn:p_Di_final_revised}
\prob(D \mid \theta_{\pop}=\Theta)= \displaystyle{\prod_{i=1}^N  }  \prob(\rmsi) \, \prob(\derri) \, \prob(d_i)  \, \prob(\pop) \times
\begin{cases} 
      \Theta \, \prob(L_{i} \geq L_{\mathrm{thr, i}} \mid \theta_{\pop}=\Theta, \pop, \rmsi,\derri, d_i, e=1) 	 	&, \,\,   \detect_i = 1 \\
      1-\Theta \,  \prob(L_{i} \geq L_{\mathrm{thr, i}} \mid \theta_{\pop}=\Theta, \pop, \rmsi,\derri, d_i, e=1)      &, \,\, \detect_i = 0 \quad.
\end{cases}
\end{equation}
Here, we have also added the multiplicative term $\prob(\pop)$, or the probability that an object is in a given population $p$, where $\pop \in [\mathrm{single}, \mathrm{binary}]$.  This allows us to condition on the appropriate luminosity distribution for each population.  The probability that an emitting object's radio luminosity is detectable now depends on the assumed binary radio occurrence rate, and by implication, the single-object radio occurrence rate.


\subsection{Constructing a predicted binary occurrence rate $\theta_{\predict}$ from the single-object occurrence rate $\theta_{\si}$ } \label{sec:changevar}

Our experiment design is straightforward: We calculate the actual quiescent emissions occurrence rate for binaries $\theta_{\bi}$  from equation \ref{eqn:p_Di_final_revised} and compare it to a ``control" occurrence rate. We design the control to reflect the scenario where binarity does \textit{not} impact the occurrence rate of quiescent emissions in ultracool dwarfs.  If the actual occurrence rate differs from the control occurrence rate, we conclude that binarity does in fact impact the occurrence rate for quiescent emissions in binaries.

To construct the control occurrence rate, we assume that binarity does not affect the radio occurrence rate of individual components or their luminosities.  In \S \ref{sec:lumDist}, we examine this latter assumption.  With these assumptions in hand, we can now calculate the occurrence rate  of quiescent emissions predicted for binaries $\theta_{\predict}$  from  the single-object occurrence rate  $\theta_{\si}$ by considering the appropriate change of variables.

The sum of both components in a binary fall into the binary luminosity range of interest such that
\begin{equation} \label{theta_binary}
    \theta_{\bi} = \prob( (L_a + L_b \geq L_{\mathrm{min}}) \cap (L_a + L_b \leq 2L_{\mathrm{max}})) \quad.
\end{equation}
We choose an appropriate $L_{\mathrm{max}}$ such that the single-object luminosity prior takes the form
\begin{align}
    \label{eqn:assumption}
    \PDF(L_{\si}) =
    \begin{cases}
        0  \quad \quad &, \quad L_{\si} > L_{\text{max}} \\ 
        \theta_{\si} \, \mathcal{F}_1(L_{\si})        \quad \quad & ,\quad L_{\si} \in [L_{\text{min}}, L_{\text{max}}] \\
        (1-\theta_{\si}) \, \mathcal{F}_2(L_{\si})    \quad \quad &, \quad L_{\si} < L_{\text{min}} \quad,
    \end{cases}
\end{align}
where we discuss $L_{\text{max}}$ in \S \ref{sec:lum_prior}, and  $\mathcal{F}_1(L_{\si})$ and  $\mathcal{F}_2(L_{\si})$ are normalized density functions that depend on $L_{\si}$. This generalized form meets the condition defined in equation \ref{eqn:probability_mass} and gives us the freedom to examine how distributions with different shapes impact the final calculated binary occurrence rate. For instance, a situation where objects  are either ``on" or ``off" can be described by a luminosity prior where the probability mass for $L_{\si} < L_{\text{min}}$ is concentrated as a delta function at $L_{\si} = 0$.  While this is not the case for thermal emission, it may be the case for non-thermal magnetospheric radio emission: not all planets in the Solar System host radiation belts, and not all ultracool dwarfs may flare.  By selecting an appropriate $L_{\text{max}}$ (\S \ref{sec:lum_prior}), equation 6 simplifies to  $\theta_{\bi} = \prob( L_a + L_b \geq L_{\mathrm{min}})$.  

We then account for all combinations of $L_a + L_b$ by invoking the law of total probability and integrating 
\begin{align} \label{eqn:numintegrate}
   \theta_{\predict} & = \int_{0}^{L_{\mathrm{max}}} \prob(L_a + k \geq L_{\mathrm{min}}) \PDF(L_b=k) \, dk 
\end{align}
where $\PDF(L_b=k)$ is the probability density distribution that the $b$ component has a luminosity equal to $k$.   This distribution is our chosen single-object luminosity distribution, such that $\PDF(L_b=k) = \PDF(L_{\si}=k)$, and it gives the relationship between $\theta_{\si}$ and $\theta_{\predict}$. Note that integrating equation \ref{eqn:binary_luminosity} for $L_{\bi} >  L_{\text{min}}$ equivalently arrives at equation \ref{eqn:numintegrate}. In Appendix \ref{sec:proof}, we show that we can write the predicted binary quiescent emissions occurrence rate as
\begin{equation} 	\label{eqn:binaryOccurrenceRate}
	\theta_{\text{\predict}} = \mathcal{C}(\theta_{\si}) + \big[  1- (1-\theta_{\si})^2 \big] \quad,
\end{equation}
where $\mathcal{C}(\theta_{\si}) = \mathbb{C}(1-\theta_{\si})$ is a correction function and $\mathbb{C}$ is a constant that depends on how the probability mass for single-object luminosities is distributed for $L_{\si} < L_{\mathrm{min}}$.   In the case where objects are either ``on" or ``off", the non-emitting probability mass is a delta function at zero and $\mathcal{C}(\theta_{\si})=0$. Intuitively, this is because the summed luminosities of an emitting and non-emitting component still fall below $L_{\mathrm{min}}$.   The occurrence rate of binaries with quiescent luminosities within $[L_{\mathrm{min}}, L_{\mathrm{max}}]$  is then
 one minus the probability that neither component is emitting.  When the mass is not distributed as a delta function, $\mathcal{C}(\theta_{\si})$ is positive and accounts for the additional pairs of objects that may be too faint individually to fall above $L_{\si} \geq L_{\mathrm{min}}$  but together are luminous enough to fall above $L_{\bi} \geq L_{\mathrm{min}}$.  From equation \ref{eqn:binaryOccurrenceRate}, we see that $\theta_{\predict} > \theta_{\si}$ and expect that unresolved binaries are more likely to emit radio emissions than their single counterparts when all else is equal.

\begin{figure*}
	\begin{centering}
		\includegraphics[width=\columnwidth]{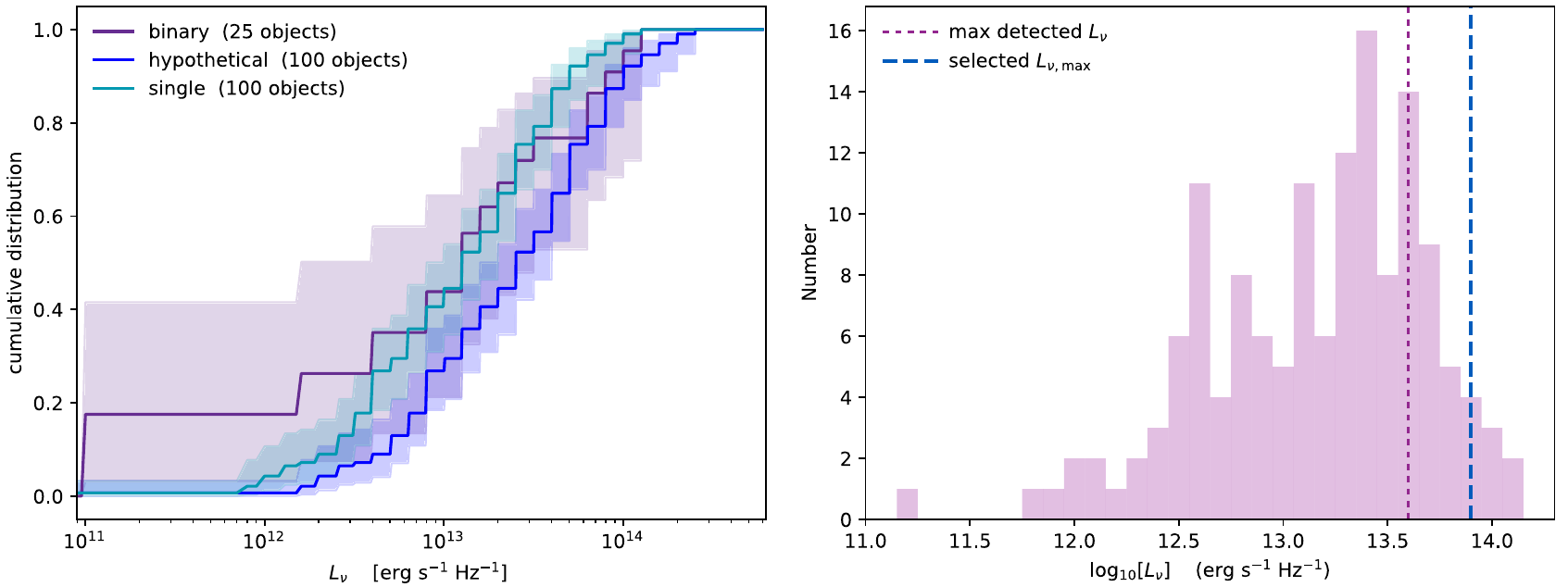}
	\end{centering}
   \caption{\textbf{Left:} Empirical cumulative distribution function for all observed single objects that meet the data inclusion criteria described in \S \ref{sec:Data} (teal), a hypothetical set of binaries that have $L_{\nu} = 2L_{\si}$ (blue), and detected binary systems (purple) within 35 pc. Shaded regions are 95 per cent confidence intervals, calculated with the exponential Greenwood's formula, a standard survival analysis error estimator \citep{FeigelsonAstroStats}. Detected objects with more than one measurement are represented with their median value. Distributions and confidence intervals are the mean distributions from 5000 trials in which we randomly draw single objects to match the spectral type distribution of the binary systems.  The low luminosity end of the binary distribution is consistent with the single-object distribution, and the high luminosity end is consistent with the hypothetical distribution. \textbf{Right:} Histogram of 4$\sigma$ luminosity limits for undetected single objects with $3\sigma$ distances within 35 pc.  85 per cent of single-object observations had luminosity limits that would have been sufficient to detect the brightest quiescent radio emissions observed in single objects (maroon dashed line). The blue dashed line shows our selected $L_{\text{max}}$, which we discuss in \S \ref{sec:lum_prior}. }
	\label{fig:ECDF_all} 
\end{figure*}

\section{Defining the luminosity distribution $\PDF(L_{\si})$} \label{sec:lumDist}

\subsection{Do individual binary components follow the same luminosity distribution as single objects?}  \label{sec:lumDistcompare} 
Self-consistent comparisons between single and binary systems require a binary luminosity prior constructed from that of single objects (\S \ref{sec.Occurrence_Rate}). For our analysis, we do not condition the underlying single-object luminosity distribution $\PDF(L_{\si})$ on binarity. One consequence of our treatment is that we expect unresolved binary systems to be more luminous on average than single objects when all else is equal.  Indeed, the data seem to support this assumption, with mean quiescent radio luminosities for detected binaries brighter than that of detected single objects \citep{kao2022ApJ...932...21K}.

Figure \ref{fig:ECDF_all} further assesses our assumption by examining empirical cumulative distribution functions (ECDFs) using the Kaplan-Meier estimator for observed single objects, binaries, and a hypothetical set of binaries constructed with our single-object sample. \S \ref{sec:Data} describes our data.  Briefly, the Kaplan-Meier estimator is a non-parametric statistic used in survival analysis \citep{Kaplan-Meier1958, Lee_survival, FeigelsonAstroStats}, which can incorporate non-detections and limits within cumulative distribution functions. In our case, it estimates the probability  that the observed object's luminosity is less than some value. 

We use the full set of compiled observations from the literature for objects that are within 35 pc, including non-detections.  This distance cutoff is consistent with all detected objects, and it includes 154 and 25 single and binary systems, respectively, or 86 per cent and 93 per cent of observed systems.  Figure \ref{fig:ECDF_all} 
shows a histogram of 4$\sigma$ sensitivity limits for all undetected single object systems with radio observations that have 3$\sigma$ distances within 35 pc.  For objects with repeated observations, we use the median measured luminosity.  We treat binaries as unresolved systems by combining the reported luminosities of each binary component for resolved systems such as 2MASS J07464256+2000321AB 
\citep{Zhang2020ApJ...897...11Z}.  Finally, each undetected object is represented only once as an upper limit corresponding to the most sensitive observation that is available in the literature for that object. 

We account for upper limits by treating non-detections as left-censored data, allowing for the possibility that non-detections are too faint to be detected. The Kaplan-Meier estimator makes the fundamental assumption that all objects in the given data set are emitting quiescent emissions, even if they are not detectable. It is important to note that this assumption may not be true, since some undetected objects may not be emitting non-thermal radio emissions at all. For instance, \citet{kao2024MNRAS.527.6835K} find that $\sim$15--20 per cent of objects may not be emitting non-thermal quiescent radio emissions brighter than $10^{11.7}$~erg~s$^{-1}$~Hz$^{-1}$, and future observations at higher sensitivities may yield a significant number of upper limits at lower luminosities.  This would shift the probability mass in the ECDFs for single and hypothetical populations leftward to lower luminosities that better align with ``on''/``off'' cases.

For our hypothetical set of binaries, we randomly draw objects from the single-object sample to match the spectral type distribution of the individual components in the binary sample.  We repeat this trial 5000 times and show the mean distribution and 95 per cent confidence interval in Figure \ref{fig:ECDF_all}.  We find that the low luminosity end of the binary distribution is consistent with the single-object distribution and the high luminosity end is consistent with the hypothetical binary distribution. Even though \citet{kao2022ApJ...932...21K} observed that the highest luminosities observed in single ultracool dwarfs could not account for the most luminous detected binaries, controlling for spectral type distributions seems to show that single and binary objects do not have statistically different underlying luminosity functions. This further grounds our choice not to condition the underlying single-object luminosity distribution $\PDF(L_{\si})$ on binarity.

\subsection{Choosing a self-consistent luminosity distribution} \label{sec:lum_prior}

To assess if binaries have elevated radio occurrence rates relative to \textit{analogous} single objects, we must use a luminosity distribution that is consistent for both the single and binary populations. \citet{kao2024MNRAS.527.6835K} adopted a luminosity prior with minimum and maximum single-object luminosities that correspond to those observed for the known population of radio-bright ultracool dwarfs at our frequencies of interest.  Although we show in  \S \ref{sec:lumDistcompare}  that the underlying luminosity distribution for the individual components of binary systems is statistically indistinguishable from that of observed objects, the actual set of detected binaries includes systems that can be more luminous than summed luminosities from the most luminous detected single objects  \citep{kao2022ApJ...932...21K}. Thus, we cannot adopt the same minimum and maximum observed single-object luminosities that \citet{kao2024MNRAS.527.6835K} used. Instead, we must revise the luminosity range.

First, we choose $L_{\text{max}}$ such that all empirical detections of single objects have $L_{\si} \leq L_{\text{max}}$ and it reasonably accounts for detection limits. Twice the brightest reported quiescent radio luminosity for single objects corresponds to $L_{\nu} \sim 10^{13.9}$ erg s$^{-1}$ Hz$^{-1}$, but the M7+M7 binary 2MASS J13142039+1320011	 (also known as NLTT 33370 AB)  has luminosities up to $L_{\nu} \sim 10^{14.6}$ erg s$^{-1}$ Hz$^{-1}$ \citep{McLean2011ApJ...741...27M}. Surprisingly, resolved observations demonstrate that only one component of this system is emitting \citep{Forbrich2016ApJ...827...22F}.  As \citet{kao2022ApJ...932...21K} note, this suggests the possibility that individual components in binary systems may be overluminous compared to single-object systems.   However, radio luminosities for this system can differ by greater than a factor of two \citep{Williams2015ApJ...799..192W, Forbrich2016ApJ...827...22F}, and some portion of its radio emissions have been attributed to gyrosynchrotron flares \citep{Williams2015ApJ...799..192W}. 

Since the radio luminosity of 2MASS J13142039+1320011 may not trace solely quiescent radio emission, we turn to the next most luminous binary system, 2MASS J13153094-2649513AB.  This L3.5+T7 binary has $L_{\nu} \sim 10^{14.2}$ erg s$^{-1}$ Hz$^{-1}$ \citep{Burgasser2013ApJ...762L...3B}, which is still overly luminous compared to the brightest single objects.  

This corresponds to a single-object maximum luminosity $L_{\text{max}} = 10^{13.9}$~erg s$^{-1}$ Hz$^{-1}$, which we choose for this calculation. This upper limit, revised from $10^{13.6}$~erg s$^{-1}$ Hz$^{-1}$, is consistent with reported binary luminosities and it includes  97\% of luminosity limits for single objects within 35 pc. Finally, it corresponds to two times the maximum quiescent radio luminosity of detected single ultracool dwarfs, which is consistent with observed levels of variability in ultracool dwarf quiescent radio emissions \citep[][and references therein]{kao2024MNRAS.527.6835K}.  The one outlier is the L2.5 dwarf 2MASS J05233822-1403022, which \citet{Antonova2007AA...472..257A} report can vary by a factor of $\sim$5 from $\leq$45 to 230$\pm$17 $\mu$Jy with no evidence of short-duration flares during 2-hr observing blocks. Although the low circular polarization of this object rules out coherent aurorae, non-auroral  flares at both radio and optical frequencies can persist for at least several hours \citep[e.g.][]{Villadsen2019ApJ...871..214V, Paudel2018ApJ...861...76P}. This outlier object has been excluded from the single-object sample by \citet{kao2024MNRAS.527.6835K} on the basis of the uncertain nature of its radio emission, similar to our reasoning for excluding the binary 2MASS J09522188-1924319 from the binary sample (see \S \ref{sec:Data}).

\begin{figure*}
	\includegraphics[width=0.6\textwidth]{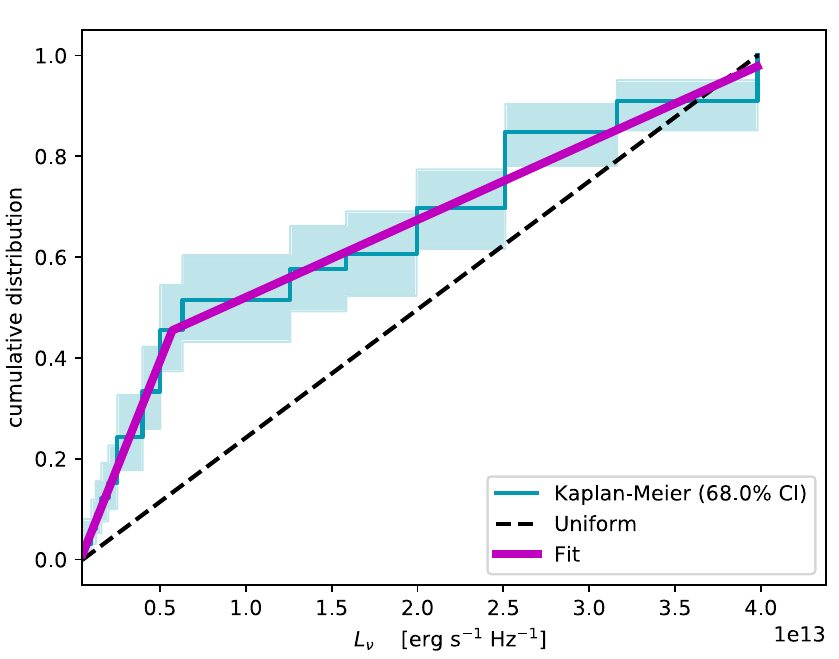}
    \caption{The empirical cumulative distribution function for $L_{\nu}$ for all radio detections of single ultracool dwarfs, calculated using the Kaplan-Meier estimator (teal). We obtain a probability density distribution calculated by differentiating this cumulative distribution function and fit a piece-wise linear prior $\PDF(L_{\si} \mid e=1)$  (magenta) as well as a uniform prior (black).  We run calculations for both of these fitted distributions.}
    \label{fig:lum_pdf} 
\end{figure*}

Following \citet{kao2024MNRAS.527.6835K}, we assume two luminosity distributions for our calculations, which we show in Figure \ref{fig:lum_pdf}:
\begin{itemize}
    \item{\textbf{Uniform$_{e=1}$:}} Emitting single objects have radio luminosities that are uniformly distributed between $L_{\si} \in [L_{\text{min}} = 10^{11.7}, L_{\text{max}}=10^{13.9}]$  erg s$^{-1}$ Hz$^{-1}$.  Here, $L_{\text{min}}$ corresponds to the minimum detected radio luminosity for single ultracool dwarfs, and our chosen $L_{\text{max}}$ provides an upper bound for existing detection measurements.  By including repeat observations of the same object, we can account for the intrinsic variability of individual objects.  Calculations that use this luminosity distribution assume that our only prior knowledge about the luminosity distribution are the minimum and maximum values, as is the case with the existing limited data.
    \item{\textbf{KM$_{e=1}$:}} Alternatively, we can assume that emitting objects follow the same luminosity distribution of \textit{all measured detections}.  We construct this prior by fitting the Kaplan-Meier ECDF of emitting single objects with a piece-wise linear function (Figure \ref{fig:lum_pdf}) and differentiating the fitted function.  This prior also accounts for intrinsic variability in individual objects, but it may be skewed from the true underlying luminosity distribution due to the non-uniformity of repeated observations and small sample size (33 measurements).  The KM probability density distribution initially extends only to $L_{\nu} = 10^{13.6}$ erg s$^{-1}$ Hz$^{-1}$, but it is characterized by two concatenated uniform distributions (Figure \ref{fig:lum_pdf}).  Accordingly, we extend the luminosity distribution to our chosen maximum limit $L_{\nu} = 10^{13.9}$ erg s$^{-1}$ Hz$^{-1}$ and re-normalize. 
\end{itemize}
Figure \ref{fig:lum_pdf} shows our chosen luminosity priors for $\PDF(L_{\si} \mid e=1)$.  Using the same luminosity priors for both the single and binary populations controls for luminosity differences between single and binary populations and allows us to ask whether binaries have elevated occurrence rates relative to analogous single objects. 
 
As discussed in \S \ref{sec.Occurrence_Rate}, the probability mass outside of the luminosity prior for emitting single objects ($e=1$) impacts the probability distribution for the binary luminosities. The single-object radio occurrence rate $\theta_{\si}$ corresponds to the probability that an object is emitting within the assumed luminosity bounds  $L_{\si} \in [10^{11.7}, 10^{13.9}]$ erg s$^{-1}$ Hz$^{-1}$.  Thus, we attribute  $\theta_{\si}$ of the probability mass for $\PDF(L_{\si})$ within these assumed bounds. Finally, we must decide how to distribute the rest of the probability mass $(1- \theta_{\si})$ between $L_{\si} \in [0, 10^{11.7})$ erg s$^{-1}$ Hz$^{-1}$.

For this work, we examine the case where $(1- \theta)$ of the mass is located at $L_{\si} = 0$.  This simplest case equivalently assumes that single ultracool dwarfs only emit radio emissions between $L_{\si} \in [10^{11.7}, 10^{13.9}]$ erg s$^{-1}$ Hz$^{-1}$.   

Our assumption is motivated by the fact that a number of factors can contribute to a lack of quiescent radio emissions from a given ultracool dwarf.  Such factors include but are not limited to the absence of a plasma source, magnetic fields that may not be sufficient in size or strength to confine the plasma, or  the absence of a mechanism that can accelerate magnetospheric plasma to the  mildly relativistic energies required for gyrosynchrotron emissions.  Indeed, planets demonstrate that not all objects meet the necessary conditions for producing quiescent radio emissions.  For example,  the magnetized planets in our solar system (Mercury, Earth, Jupiter, Saturn, Uranus, Neptune) are ``on", whereas the unmagnetized solar system planets (Mars, Venus) are ``off" \citep[]{khurana2004, bolton2004, girard2016, MaukFox2010JGRA..11512220M, kellerman1970, basharinov1974, ganushkina2011}. Similarly, deep observations of the nearest brown dwarf binary system Luhman 16AB \citep[2.02$\pm$0.15 pc,][]{Luhman2013ApJ...767L...1L} and the Y dwarf WISE J085510.83-071442.5 \citep[2.23$\pm$0.16 pc,][]{Tinney2014ApJ...796...39T} rule out radio emissions down to $L_{\nu} \leq 10^{11.0}$ and $\leq 10^{11.2}$ erg s$^{-1}$ Hz$^{-1}$, respectively, at 4$\sigma$ significance \citep{Osten2015ApJ...805L...3O, Kao2019MNRAS.487.1994K}.

Assuming that ultracool dwarfs emit only within the bounds $L_{\si} \in [10^{11.7}, 10^{13.9}]$ erg s$^{-1}$ Hz$^{-1}$ captures the observed range of detected ultracool dwarf quiescent radio luminosities. However, the existing set of detected objects may not fully capture the faint end of the ultracool dwarf luminosity distribution.  Thus, we also include calculations for a hypothetical $\PDF(L_{\si})$ with $L_{\si}\in [10^{9.8}, 10^{13.9}]$  erg s$^{-1}$ Hz$^{-1}$.  Existing quiescent radio luminosities span 1.9 dex, and  the revised lower limit reflects an additional 1.9 dex.

\section{Applying the framework}  \label{section:application} 


\setlength{\tabcolsep}{0.16in}
\begin{center}
	\begin{table*}\centering
		\begin{ThreePartTable}
			\caption{Radio Observations of Multi-object Ultracool Dwarf Systems from the Literature  \label{table:binariesBD}}
			\begin{tabularx}{\textwidth}{ llcclll }
			\toprule \vspace{2pt}
     Object Name            & SpT           & \multicolumn{1}{c}{$\pi$} & \multicolumn{1}{c}{$d$}       & \multicolumn{1}{c}{$F_{\nu}$} & \multicolumn{1}{c}{ref} & Note \\
			                &               & \multicolumn{1}{c}{(mas)} & \multicolumn{1}{c}{(pc)}      & \multicolumn{1}{c}{($\mu$Jy)} &                    &  \\
			  \midrule
LP 415-20                   &   M6 + M8           & 25.1963 $\pm$ 0.5117    &   39.7    $\pm$   0.8         &   24.1 $\pm$ 1.4  &	24	40	28	35	&   a $\dag$ \\
2MASS J00244419-2708242	    &	M6 + M8.5 + M9	  &	132.3	$\pm$	11.4	&	7.6	    $\pm$	0.7	        &	$<$111.0		&	43	38	59	46	&	b	$\dag$\\
2MASS J00275592+2219328	    &	M7 + M8	          &	69.2	$\pm$	0.9	    &	14.5	$\pm$	0.2	        &	323	$\pm$ 14	&	24	24	24	49	&	c	\\
2MASS J04291842-3123568	    &	M7.5 + L1	      &	59.3757	$\pm$	0.2018	&	16.8	$\pm$	0.1        &	$<$48			&	55	56	28	6	&		\\
2MASS J09522188-1924319	    &	M7 + M7	          &	34.5193	$\pm$	0.1511	&	29.0	$\pm$	0.1        &	233	$\pm$ 15	&	47	52	28	49	& d	$\dag$	\\
2MASS J21402931+1625183     &   M8.0 + L0.5       & 30.1972 $\pm$ 0.434     &   33.1    $\pm$   0.5         &   15.6 $\pm$ 1.7  &	24	40	28	35	&       \\  
2MASS J11214924-1313084	    &	M8.5 + L7.5	      &	69.4903	$\pm$	0.1760	&	14.4	$\pm$	0.0        &	$<$102			&	3	24	28	49	&		\\
2MASS J12073346-3932539	    &	M8 + L5	          &	19.1	$\pm$	0.4	    &	52.4	$\pm$	1.1	        &	$<$29			&	31	7	23	51	&		\\
2MASS J12560183-1257276 Aab	&	M7.5 + M7.5	      &	78.8	$\pm$	6.4	    &	12.7	$\pm$	1.0	        &	60	$\pm$	3	&	29	29	29	33	&	e	\\
2MASS J17072343-0558249	    &	M9 + L3	          &            ---          &	7.0	    $\pm$	1.0	        &	$<$48	        &	53	48	26	6	&		\\
2MASS J22000201-3038327	    &	M8 + L0	          &	41	    $\pm$	4	    &	24.4	$\pm$	2.4	        &	$<$78			&	26	12	44	49	&		\\
2MASS J22062280-2047058	    &	M8.0 + M8.5	      &	35.8	$\pm$	1.0	    &	27.9	$\pm$	0.8	        &	$<$84			&	20	24	24	49	&		\\
GJ 569 Bab	                &	M8.5 + M9         &	93.814	$\pm$	0.724	&	10.7	$\pm$	0.1	        &	$<$30			&	39	24	28	6	&	f $\dag$	\\
WISE J072003.20-084651.2	&	M9 + T5           & 147.1$^{+1.1}_{-1.2}$   &	6.8		$_{-0.06}^{+0.05}$	&	15	$\pm$	3	&	18	25	25	19	&		\\
GJ 564 BC                   &   L4 + L4           & 54.9068 $\pm$ 0.0684    &   18.2    $\pm$ 0.0           &   31.3 $\pm$ 2.2  &	24	40	28	35	&   g $\dag$     \\
2MASS J00043484-4044058	    &	L5 + L5	          &	82.0946 $\pm$   0.3768	&	12.2	$\pm$	0.1	        &	100	$\pm$	8.3	&	53	32	28	46	&	h	\\
2MASS J02052940-1159296	    &	L5 + L8 + T0	  &	50.6	$\pm$	1.5	    &	19.8	$\pm$	0.6	        &	$<$30			&	10	10	27	6	&	\,\,\,  $\dag$	\\
2MASS J03105986+1648155	    &	L9 + L9	          &	36.9	$\pm$	3.4	    &	27.1	$\pm$	2.5	        &	$<$10.8			&	36	58	57	54	&		\\
2MASS J04234858-0414035	    &	L6.5 + T2	      &	67.8584	$\pm$	1.5052	&	14.7	$\pm$	0.3	        &	54.1 $\pm$	2.2	&	23	24	28	34	&		\\
2MASS J07003664+3157266	    &	L3 + L6.5	      &	88.2790	$\pm$	0.3479	&	11.3	$\pm$	0.0	        &	$<$42			&	23	24	28	1	&		\\
2MASS J07464256+2000321 A	&	L0	              &	80.9	$\pm$	0.8	    &	12.4	$\pm$	0.1	        &	71	$\pm$	12	&	9	24	24	60	&		\\
2MASS J07464256+2000321 B	&	L1.5    	      &	80.9	$\pm$	0.8	    &	12.4	$\pm$	0.1	        &	128	$\pm$	14	&	9	24	24	60	&		\\
2MASS J10491891-5319100 	&	L7.5 + T0.5       &	496	    $\pm$	37	    &	2.0	    $\pm$	0.2	        &	$<$15			&	17	42	45	50	&		\\
2MASS J12281523-1547342	    &	L5.5 + L5.5	      &	48.0	$\pm$	1.7	    &	20.8	$\pm$	0.7	        &	$<$87			&	23	21	24	5	&	i	\\
2MASS J13054019-2541059 	&	L2 + L3.5	      &	53.8492	$\pm$	0.7107	&	18.6	$\pm$	0.2	        &	$<$27.6			&	37	24	28	41	&		\\
2MASS J13153094-2649513	    &	L3.5 + T7	      &	53.8729	$\pm$	1.1265	&	18.6	$\pm$	0.4	        &	370	$\pm$	50	&	14	15	28	16	&		\\
2MASS J14413716-0945590	    &	L1 + L1	          &	31.6439	$\pm$	1.0131	&	31.6	$\pm$	1.0	        &	$<$84			&	53	8	28	49	&		\\
2MASS J17114573+2232044	    &	L5.0 + T5.5	      &	33.11	$\pm$	4.81	&	30.2	$\pm$	4.4	        &	$<$11.4			&	13	4	27	54	&		\\
2MASS J17281150+3948593	    &	L5 + L6.5	      &	36.4	$\pm$	0.6	    &	27.5	$\pm$	0.5	        &	$<$54			&	30	24	24	6	&		\\
2MASS J22521073-1730134 	&	L4.5 + T3.5	      &	59.1461	$\pm$	0.8244	&	16.9	$\pm$	0.2	        &	$<$30			&	23	24	28	6	&		\\
2MASS J15344984-2952274 	&	T4.5 + T5         &	63.0	$\pm$	1.1	    &	15.9	$\pm$	0.3	        &	$<$63			&	23	24	24	6	&		\\
2MASS J22041052-5646577 A	&	T1 + T6	          &	275.79	$\pm$	0.69	&	3.63	$\pm$	0.01	    &	$<$79.2			&	11	22	45	2	&	j	\\
2MASS J22041052-5646577 B	&	T1 + T6	          &	275.79	$\pm$	0.69	&	3.63	$\pm$	0.01	    &	$<$79.2			&	11	22	45	2	&		\\
			\bottomrule
			\end{tabularx}	
			\begin{tablenotes}[]\footnotesize
				\item[] Note -- We do not include the radio-bright T dwarf binary WISEP J101905.63+652954.2 in this table since \citet{Vedantham2023A&A...675L...6V} report only bursting emission. Distances calculated from parallaxes are provided for the reader's convenience but truncated to three significant figures. We report upper limits or flux densities as presented in the literature but define detected = 1 if the signal-to-noise ratio $\geq 4.0$ to remain consistent with our occurrence rate calculation.  
                \item[$\dag$] Excluded from the binary sample (see \S\ref{sec:Data}).
                \item[a] The primary component of this system may be an unresolved ultracool dwarf binary \citep{Dupuy2017ApJS..231...15D}.
				\item[b] Observed with ATCA in 6A configuration, which cannot resolve this triple system. 
				\item[c] \citet{McLean2012ApJ...746...23M} list 2MASS J00275592+2219328 as LP 349-25 B, but the VLA cannot resolve this binary.
				\item[d] \citet{Reid2002AJ....124..519R} report that 2MASS J09522188-1924319 is an unresolved double-lined spectroscopic binary,  but \citet{Guenther2003AA...401..677G} observed only a single line. Its detected radio emission may  be  flaring  rather  than  quiescent  emission \citep{McLean2012ApJ...746...23M}. 
				\item[e] The discovery that the primary in the 2MASS J12560183-1257276 AB system is actually an equal-mass binary \citep{Stone2016ApJ...818L..12S} suggests that this system may be as far as $17.1 \pm 2.5$ pc rather than the $12.7\pm1.0$ pc reported by \citet{Gauza2015ApJ...804...96G}.  The wide $\geq 102 \pm 9$ AU separation between the primary and secondary components \citep{Gauza2015ApJ...804...96G} is easily resolved in all observations. 
				\item[f] GJ 569 Bab is a companion to an M3 primary that is separated by $\sim$53 AU.  This corresponds to $\sim$5'' \citep{Forrest1988ApJ...330L.119F}  and is resolved by the VLA. 
				\item[g] GJ~564~BC resides in a triple system, where the primary component is a solar analog at a distance of $\sim$47 AU \citep{Potter2002ApJ...567L.133P, Dupuy2009ApJ...692..729D}.  This separation is not resolvable by the VLA for the reported observations. 
                \item[h] 2MASS J00043484-4044058 (GJ 1001) is a triple system, in which the M4 primary is separated by $\sim$180 pc from the L5+L5 binary secondary \citep{Golimowski2004AJ....128.1733G}.  The observation reported here is for the binary component of the triple.  This distance is easily resolved, though \citet{Golimowski2004AJ....128.1733G} do not give error estimates for their measured angular separation.
				\item[i] \citet{Dupuy2017ApJS..231...15D} note that the orbit fit for 2MASS J12281523-1547342 is of questionable  quality.
				\item[j] 2MASS J22041052-5646577 is also known as $\epsilon$ Ind Bab.
				\item[] References -- 
                (1)	\citet{Antonova2013AA...549A.131A}~;
				(2)	\citet{Audard2005ApJ...625L..63A}~;
				(3)	\citet{BardalezGagliuffi2014ApJ...794..143B}~;
				(4)	\citet{BardalezGagliuffi2015AJ....150..163B}~;
				(5)	\citet{Berger2002ApJ...572..503B}~;
				(6)	\citet{Berger2006ApJ...648..629B}~;
				(7)	\citet{Blunt2017AJ....153..229B}~;
				(8)	\citet{Bouy2003AJ....126.1526B}~;
				(9)	\citet{Bouy2004AA...423..341B}~;
				(10)	\citet{Bouy2005AJ....129..511B}~;
				(11)	\citet{Burgasser2005AJ....129.2849B}~;
				(12)	\citet{Burgasser2006AJ....131.1007B}~;
				(13)	\citet{Burgasser2010ApJ...710.1142B}~;
				(14)	\citet{Burgasser2011ApJ...735..116B}~;
				(15)	\citet{Burgasser2011ApJ...739...49B}~;
				(16)	\citet{Burgasser2013ApJ...762L...3B}~;
				(17)	\citet{Burgasser2013ApJ...772..129B}~;
				(18)	\citet{Burgasser2015AJ....149..104B}~;
				(19)	\citet{Burgasser2015AJ....150..180B}~;
				(20)	\citet{Deshpande2012AJ....144...99D}~;
				(21)	\citet{Dieterich2014AJ....147...94D}~;
				(22)	\citet{Dieterich2018ApJ...865...28D}~;
				(23)	\citet{Dupuy2012ApJS..201...19D}~;
				(24)	\citet{Dupuy2017ApJS..231...15D}~;
				(25)	\citet{Dupuy2019AJ....158..174D}~;
				(26)	\citet{Faherty2009AJ....137....1F}~;
				(27)	\citet{Faherty2012ApJ...752...56F}~;
				(28)	\citet{Gaia20182018AA...616A...1G}~;
				(29)	\citet{Gauza2015ApJ...804...96G}~;
				(30)	\citet{Gelino2014AJ....148....6G}~;
				(31)	\citet{Gizis2002ApJ...575..484G}~;
				(32)	\citet{Golimowski2004AJ....128.1733G}~;
				(33)	\citet{Guirado2018AA...610A..23G}~;
				(34)	\citet{Kao2016ApJ...818...24K}~;
                (35)    \citep{kao2022ApJ...932...21K} ;
				(36)	\citet{Kirkpatrick2000AJ....120..447K}~;
				(37)	\citet{Koen2013MNRAS.428.2824K}~;
				(38)	\citet{Kohler2012AA...541A..29K}~;
				(39)	\citet{Konopacky2010ApJ...711.1087K}~;
                (40)    \citet{Konopacky2012ApJ...750...79K}    ;
				(41)	\citet{Krishnamurthi1999AJ....118.1369K}~;
				(42)	\citet{Lazorenko2018AA...618A.111L}~;
				(43)	\citet{Leinert2000AA...353..691L}~;
				(44)	\citet{Liu2016ApJ...833...96L}~;
				(45)	\citet{Luhman2013ApJ...767L...1L}~;
				(46)	\citet{Lynch2016MNRAS.457.1224L}~;
				(47)	\citet{McCaughrean2002AA...390L..27M}~;
				(48)	\citet{McElwain2006AJ....132.2074M}~;
				(49)	\citet{McLean2012ApJ...746...23M}~;
				(50)	\citet{Osten2015ApJ...805L...3O}~;
				(51)	\citet{OstenJayawardhana2006ApJ...644L..67O}~;
				(52)	\citet{Reid2002AJ....124..519R}~;
				(53)	\citet{Reid2008AJ....136.1290R}~;
				(54)	\citet{Richey-Yowell2020}~;   
				(55)	\citet{Schmidt2007AJ....133.2258S}~;
				(56)	\citet{Siegler2005ApJ...621.1023S}~;
				(57)	\citet{Smart2013MNRAS.433.2054S}~;
				(58)	\citet{Stumpf2010AA...516A..37S}~;
				(59)	\citet{Weinberger2016AJ....152...24W}~;
				(60)	\citet{Zhang2020ApJ...897...11Z}	
			\end{tablenotes}
				\end{ThreePartTable}  
	\end{table*}
\end{center}

\subsection{Data} \label{sec:Data}

To compare the radio occurrence rates of binary systems to single objects, we use the compilations of single objects in \citet{kao2024MNRAS.527.6835K}.  This set includes 82 ultracool M dwarfs, 74 L dwarfs, and 23 T/Y dwarfs.  To define the luminosity priors that we assume for our occurrence rate calculations in \S \ref{sec:lum_prior}, we use the combined dataset of all detected quiescent radio emissions measurements for single objects and binaries, the latter of which are compiled in Table 4 of \citet{kao2022ApJ...932...21K}.

Finally, we performed a literature search to compile a list of all radio observations, including non-detections, of ultracool dwarf binaries with individual components that have spectral type $\geq$M7 (Table~\ref{table:binariesBD}).  For objects with multiple recorded observations, we follow the data inclusion policy described in \citet{kao2024MNRAS.527.6835K} to produce a data set that represents observations that are independent from each other: In this data set, each object is represented only once.  For each non-detected object, we select its most sensitive observation.  For each detected object, we select the detection with the lowest rms noise. Furthermore, the flux densities that we list correspond to quiescent emissions and exclude contributions from identified flares. This latter dataset serves as the input dataset for our occurrence rate calculations in \S \ref{section:application}.

Our binary sample contains 27 systems, summarized in Table \ref{table:binariesBD}.  For binaries with reported radio emissions for separate components, we choose the higher rms value to be conservative and designate the binary as detected if at least one component was detected.  We include only binary systems for which both components are ultracool dwarfs and exclude triples except for hierarchical systems in which observations resolve the binary and where we can reasonably treat the binary as a self-contained system for our science case.

The triple system 2MASS J12560183-1257276 consists of an L7.0 secondary that is widely separated at $\geq 102 \pm 9$ AU from its primary component, an M7.5+M7.5 close-in binary system \citep{Gauza2015ApJ...804...96G, Stone2016ApJ...818L..12S}. We include the primary component because it is easily resolved from the secondary in all observations. In units of host star stellar radii, this large separation is $\gtrsim$300 times that between the Sun and Neptune \citep{pineda2021ApJ...918...40P}, the latter of which has a faint radiation belt populated by the Solar wind \citep{MaukFox2010JGRA..11512220M}.  By mass conservation, stellar wind densities approximately drop off as $1/a^2$ at far separations $a$, so we include the secondary in the single-object sample.

We also include the binary systems 2MASS J00043484-4044058 BC (GJ 1001 BC) and 2MASS J22041052-5646577 ($\epsilon$ Ind Bab) using similar reasoning.  These systems are  part of  widely-separated triple systems in which the primary components are separated from the secondary binary systems by $\sim$180 AU \citep[18\arcsec,][]{Golimowski2004AJ....128.1733G} and $\sim$1459 AU \citep[402\arcsec,][]{Scholz2003AA...398L..29S}.  
We do not include the primaries for these systems in the single sample because they are not ultracool dwarfs. Instead, they are M4 and K5V stars, respectively.

We exclude from our binary sample 2MASS~J00244419-2708242, LP~415-20,  GJ~564~BC, GJ~569~Bab, and 2MASS J09522188-1924319.  For 2MASS~J00244419-2708242, the ATCA 6A configuration cannot resolve this triple system.  For LP~415-20,  compelling evidence suggests that the primary component of this system, which radio observations cannot resolve from its secondary, may in fact itself be an unresolved ultracool dwarf binary system \citep{Dupuy2017ApJS..231...15D}.   GJ~564~BC resides in a triple system that is unresolved by the VLA. The primary component is a solar analog at a distance of $\sim$47 AU, or $\sim$1.6 times the separation of Neptune from the Sun \citep{Potter2002ApJ...567L.133P, Dupuy2009ApJ...692..729D}. Although solar radio emissions from 4--8 GHz during solar maximum at this object's distance is not detectable by the VLA \citep{Villadsen2014ApJ...788..112V}, we cannot rule out the possibility that the solar wind from the primary may populate radiation belts around the secondary binary, so we cannot confidently treat the secondary component as a self-contained binary system.  GJ~569~Bab is also in a triple and separated from its M3 primary by $\sim$50 AU \citep[5\arcsec,][]{Forrest1988ApJ...330L.119F}.  Although \citet{Berger2006ApJ...648..629B} do not report the observation date, VLA configuration, or observing frequency for this target, we surmise from an archive search that this target was observed during VLA B configuration at C band (Project Code AB1179, PI Bower).  The separation between GJ~569~Bab and its primary is resolvable by the VLA for this observation. However, $\sim$50 AU is only $\sim$4.5 times the stellar radii separation between Neptune and the Sun, so we cannot be confident that GJ~569~Bab remains largely uninfluenced by its primary star. Finally, we exclude the M7+M7 binary 2MASS J09522188-1924319 because the radio emissions detected from this object may be flaring rather than quiescent emissions. \citet{McLean2012ApJ...746...23M} detected $233 \pm 15$ $\mu$Jy emissions from this object, which corresponds to $L_{\nu} = 10^{14.4}$~erg~s$^{-1}$~Hz$^{-1}$. However, they reported that followup observations at 4.96 and 8.46 GHz after the initial detection did not yield a detection to a limit of 69 $\mu$Jy, or a factor of 2.4 below the original detection.  They concluded that the initial detection was likely a flare or that 2MASS J09522188-1924319 exhibits long-term variability. Excluding this object gives a more conservative binary quiescent radio occurrence rate.

\subsection{Analysis}

\begin{figure*}
	\includegraphics[width=\textwidth]{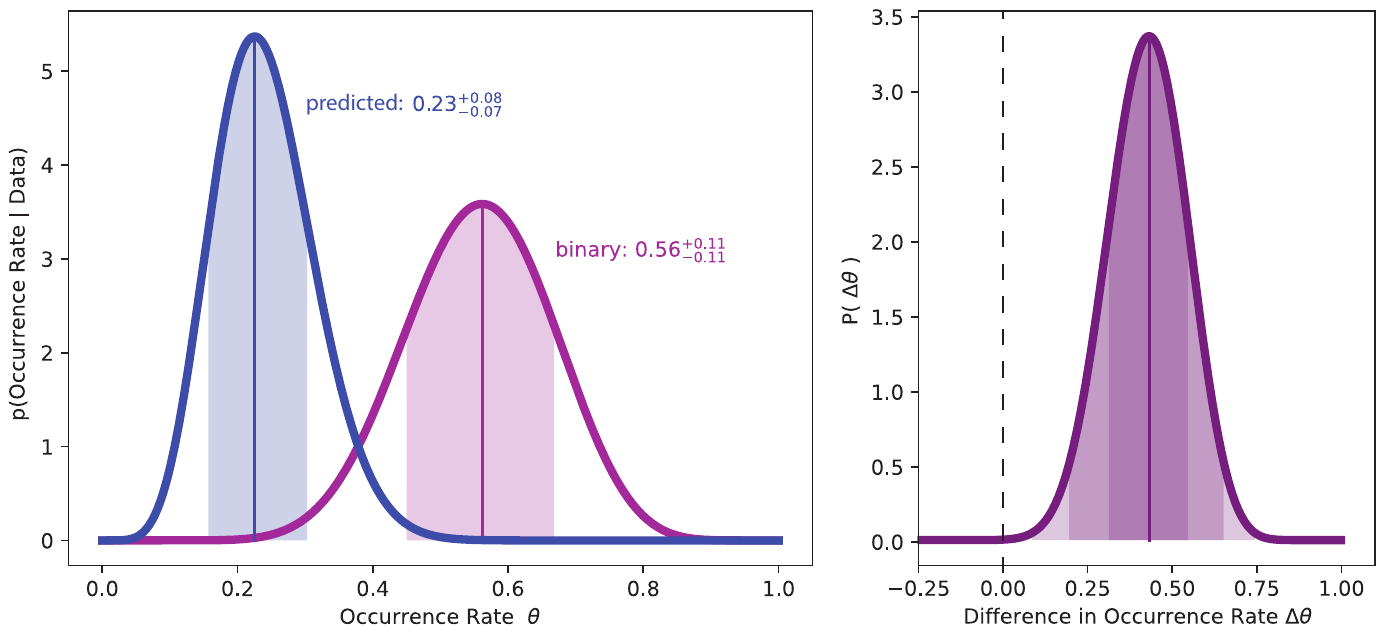}
    \caption{Calculations comparing binary to single-object systems using the full set of data available in the literature and assuming a uniform luminosity prior between $L_{\nu, \si} \in [10^{11.7}, 10^{13.9}]$~erg~s$^{-1}$~HZ$^{-1}$. \textbf{Left:} Occurrence rate distributions of quiescent radio emissions in ultracool dwarf binaries (mauve) and the predicted occurrence rate distribution for binaries from single objects, where $\theta_{\text{predicted}} = 1- (1-\theta_{\text{single}})^2$ (blue).  Shaded regions correspond to the 68.3\% credible intervals. Individual components in binary systems are more likely to emit radio emissions than their single counterparts. \textbf{Right:} Probability density distribution for the difference in occurrence rates $\Delta \theta = \theta_{\bi} - \theta_{\predict}$ between binaries and predictions from single objects.  Shaded regions correspond to 68.3\%, 95.5\% and 99.7\% credible intervals.  Existing data rule out the null hypothesis that binarity does not affect quiescent radio occurrence rates.   }
	\label{fig:occurrenceRateBinary}
\end{figure*}

To assess if binaries have an observed quiescent radio occurrence rate that is consistent with that of analogous single-objects, we compare occurrence rate distributions for two populations in Figure \ref{fig:occurrenceRateBinary}: the set of observed binaries, and a control set of constructed binaries. 

For the control set, we randomly draw single-object systems without replacement to match the M, L, and T dwarf distribution of the individual components in the binary sample.  The number of observed T dwarf systems limits the sizes of our drawn samples to 108 objects.  We account for different possible combinations of single-object systems by repeating this procedure 5000 times and calculating the probability density distribution of the occurrence rate for each trial using the single-object framework defined in \citet{kao2024MNRAS.527.6835K}. In other words, this occurrence rate reflects the occurrence rate that one would observe if each binary were split into its individual components.

We then take the mean distribution for single objects $\PDF(\theta_{\text{single}})$ and transform it to a \textit{predicted} distribution for our hypothetical binaries $\PDF(\theta_{\predict})$ by ensuring that 
\begin{equation}
{\displaystyle\int\limits_{\theta_{\si,1}}^{\theta_{\si,2}}} \PDF(\theta_{\si}) d\theta_{\si}  =  {\displaystyle \int\limits_{\theta_{\predict,1}}^{\theta_{\predict,2}}}\PDF(\theta_{\predict}) d\theta_{\predict} \quad,
\end{equation}
which is the standard transformation for random variables. Equation \ref{eqn:binaryOccurrenceRate}  gives $\theta_{\predict} = 1- (1-\theta_{\si})^2$. 
We compare $\PDF(\theta_{\text{predicted}})$ to the actual calculated binary radio occurrence rate distribution $\PDF(\theta_{\text{binary}})$  in Figure \ref{fig:occurrenceRateBinary}.  

Finally, we also calculate the probability density distribution for differences in occurrence rates $\Delta \theta$ between predicted and actual binary systems, where
\begin{equation}
\PDF(\Delta \theta) = \int_{-1}^{1} \PDF(\theta_{\predict}) \PDF(\theta_{\bi} = \theta_{\predict} + \Delta\theta) \, d \theta_{\predict}  \quad.
\end{equation} 
We can then calculate the probability that the binary radio occurrence rate exceeds the predicted occurrence rate by integrating
\begin{equation}
    \prob(\theta_{\bi} > \theta_{\predict}) = \int_{0}^{1} \PDF(\Delta \theta) \, d \Delta\theta   \quad.
\end{equation}

\subsection{Results: Occurrence rate of binary systems compared to predictions from single objects } \label{sec:results_occurrenceRate}

In  Figure \ref{fig:occurrenceRateBinary}, we show that binary ultracool dwarf systems have an elevated radio occurrence rate compared to the rate predicted from their single counterparts.  This figure uses the full set of objects that have been observed at radio frequencies and assumes that  $L_{\nu,\si} \in [10^{11.7},10^{13.9}]$ erg s$^{-1}$ Hz$^{-1}$.

We find a quiescent occurrence rate of $56^{+11}_{-11}$ per cent for the binary sample, where uncertainties correspond to 68.3 per cent credible intervals.  In comparison, the predicted binary occurrence rate from single-object systems with the same spectral type distribution as our binary systems is $23^{+8}_{-7}$ per cent. The radio occurrence rate of binaries exceeds the predicted rate from single objects with probability $\prob(\theta_{\bi} > \theta_{\predict}) = 100$~per~cent, so we conclusively rule out the null hypothesis and further explore implications for the observed binary enhancement of radio occurrence rates in \S \ref{sec.Discussion}. 
 
We also find that binarity may enhance the ultracool dwarf quiescent radio occurrence rate even when we exclude objects from \citet{Kao2016ApJ...818...24K}.  This targeted study selected objects with likely markers of auroral magnetic activity at other wavelengths such as H$\alpha$ emissions and infrared variability.  These selection criteria resulted in a very high detection rate (80 per cent) compared to the 5--10 per cent detection rate in the literature \citep[e.g.][]{RouteWolszczan2016ApJ...830...85R, Richey-Yowell2020} that is consistent with that of volume-limited surveys \citep[e.g.]{Antonova2013AA...549A.131A, Lynch2016MNRAS.457.1224L}. While \citet{Richey-Yowell2020} showed that infrared photometric variability at 0.5--4 $\mu$m does not trace quiescent radio emissions, but they reaffirm an earlier finding by \citet{Pineda2017ApJ...846...75P} that  H$\alpha$ emissions are correlated with the quiescent radio  of ultracool dwarfs that have detected radio aurorae.  Excluding these objects removes any obvious bias for magnetic activity in our samples. With the exception of the \citet{Kao2016ApJ...818...24K} study, other survey attempts to introduce a positive detection bias have not yielded detection rates that are distinct from volume-limited studies \citep[e.g.][]{Richey-Yowell2020}.  Similarly, attempts to target individual objects that may be promising have yielded mixed results \citep[e.g.][]{Audard2005ApJ...625L..63A, Pineda2018ApJ...866..155P}.  We therefore elect to include all remaining available data to allow the hypothetical effects of these various selection attempts to average out.

For both datasets, we repeat this calculation for four luminosity distribution cases. We summarize these calculations in Table \ref{table:calculations}.  In all cases, binary ultracool dwarfs have an elevated quiescent radio occurrence rate. 

The detection rate for our sample is $37^{+11}_{-11}$ per cent (with bootstrapped 68.3 per cent confidence intervals), or $\sim$57-69 per cent of the calculated binary quiescent radio occurrence rate depending on the luminosity prior. This is consistent with correction factors between $\sim$51-66 per cent calculated for single-object systems using simulated data that followed literature distributions for relevant properties \citep{kao2024MNRAS.527.6835K}. These correction factors can be attributed largely to noise floors, such that detection rates begin to converge with occurrence rates as rms noise for samples decrease \citep{kao2024MNRAS.527.6835K}. This calls for re-observing undetected systems with high noise floors, which may yield new detections.

\setlength{\tabcolsep}{0.1in}
\begin{center}
	\begin{table*}\centering 
		\begin{ThreePartTable}
			\caption{Calculated occurrence rates \label{table:calculations}}
			\begin{tabularx}{0.75\textwidth}{ lllrr@{\hspace{0.01in}}lr@{\hspace{0.01in}}lc }
				\toprule \vspace{2pt}
				Dataset & Sample Size & $\PDF(L_{\nu, \si})$  & $[L_{\min}], [L_{\max}]$ &  \multicolumn{2}{c}{$\theta_{\bi}$ } &  \multicolumn{2}{c}{$\theta_{\predict}$ } & $\prob(\theta_{\bi}  > \theta_{\predict})$  \\	
				& (binary, single) &\multicolumn{1}{c}{$ e = 1$} & \multicolumn{1}{c}{$ e = 1$} & \multicolumn{2}{c}{ } & \multicolumn{2}{c}{ } & \\
				\midrule 
				Literature				& 27, 108	&   Uniform	&   11.7--13.9	&   0.56 &  $_{-0.11}^{+0.11}$	&   0.23 & $_{-0.07}^{+0.08}$	&   100\%  	\\[2pt]
				Literature				& 27, 108	&   Uniform	&   9.8--13.9	&   0.57 &  $_{-0.11}^{+0.11}$	&   0.23 & $_{-0.07}^{+0.08}$	&   100\%  	\\[6pt]
				Literature  	        & 27, 108	&   KM  	&	11.7--13.9  &   0.64 &	$_{-0.12}^{+0.11}$  &   0.27 & $_{-0.08}^{+0.09}$	&   100\% 	\\[2pt]
				Literature  			& 27, 108	&   KM  	&	9.8--13.9   &   0.65 &	$_{-0.12}^{+0.11}$  &   0.28 & $_{-0.08}^{+0.09}$	&   100\% 	\\[6pt]
				
                no K16 $^{\tiny{a}}$  	& 26, 52	&   Uniform	&	11.7--13.9  &   0.54 &  $_{-0.11}^{+0.11}$ 	&   0.16 & $_{-0.09}^{+0.12}$	&   100\%  	\\[2pt]
				no K16 $^{\tiny{a}}$  	& 26, 52	&   Uniform	&	9.8--13.9   &   0.54 &  $_{-0.11}^{+0.11}$ 	&   0.16 & $_{-0.09}^{+0.12}$	&   100\%  	\\[6pt]
				no K16 $^{\tiny{a}}$ 	& 26, 52	&   KM		&	11.7--13.9  &   0.61 &	$_{-0.12}^{+0.12}$  &   0.20 & $_{-0.11}^{+0.15}$	&   100\% 	\\[2pt]
				no K16 $^{\tiny{a}}$ 	& 26, 52	&   KM		&	9.8--13.9   &   0.63 &	$_{-0.12}^{+0.12}$  &   0.20 & $_{-0.12}^{+0.15}$	&   100\% 	\\[2pt]
				\bottomrule
			\end{tabularx}	
				
			\begin{tablenotes}[]\footnotesize
				\item[]Note -- For the single sample, single objects were randomly drawn from a samples of 82 ultracool M dwarfs, 74 L dwarfs, and 23 T/Y dwarfs to match the spectral type distribution of the individual components in the binary sample. The number of observed T dwarf systems limits sample sizes for constructed single-object samples.  Listed values are calculated from the mean distribution of 5000 trials.  Reported uncertainties correspond to 68.3\% credible intervals. 
				\item[$a$]Excluding observations from \citet{Kao2016ApJ...818...24K,Kao2018ApJS..237...25K} to reduce possible bias from their selection effects.  
			\end{tablenotes}
		\end{ThreePartTable}  
	\end{table*}
\end{center}

\section{Discussion}\label{sec.Discussion}

Comparing binary versus  single-object systems gives strong evidence that binarity may enhance the occurrence rate of quiescent radio activity in ultracool dwarfs.  The radio occurrence rates that we report are specifically for non-flaring non-thermal radio emissions, suggesting that radiation belts are more likely to develop around ultracool dwarfs in binary systems than in their single-object counterparts. We additionally rule out excess flaring induced by magnetic interactions \citep[e.g.][]{Morgan2012AJ....144...93M,Lanza2012A&A...544A..23L}. The smallest separation binaries in our sample ($\sim0.6$~AU) exclude significant magnetospheric interactions. The strongest ultracool dwarf dipole fields \citep[$\sim$5 kG;][]{Kao2018ApJS..237...25K} will decay as $1/r^3$ to  $\lesssim20$ nG at the half-way point between binary components, which is less than ISM field strengths \citep{Sofue2019MNRAS.485..924S}.  

Since the quiescent emissions that we focus on does not directly trace flares \citep[e.g.][see also \S\ref{sec:intro} and \S\ref{sec:Data} in this paper]{Kao2023Natur.619..272K}, our finding of a binary enhancement effect raises interesting questions about the plasma sources for this emissions.  

One possibility is that volcanic activity from satellites may not be the dominant source of emitting magnetospheric plasma for ultracool dwarfs. Planet occurrence rates are suppressed for close-in binary stellar systems compared to wide-separation binaries and single-star systems  \citep{Moe2019arXiv191201699M}. Theory suggests that this suppression occurs because binaries can truncate the mass and radius of the circumprimary disk, increase disk turbulence, and/or clear out disk material on timescales faster than planet formation \citep[e.g.,][]{Artymowicz1994ApJ...421..651A, Haghighipour2007ApJ...666..436H,Rafikov2015ApJ...798...69R}.  Like stars, ultracool dwarfs are known to host planets \citep[e.g. the TRAPPIST-1 system,][]{Gillon2017Natur.542..456G, hardegree-ullman2019AJ....158...75H} and theory predicts multiple rocky satellites around brown dwarfs or gas giant planets \citep{CanupWard2006Natur.441..834C, He2017MNRAS.464.2687H, Cilibrasi2021MNRAS.504.5455C}.  Thus,  similar mechanisms may suppress planet formation around close-in brown dwarf binaries. In this scenario, we would also expect a suppressed quiescent radio occurrence rate if volcanic satellites are a key electron source. 

We emphasize, however, that our results do not rule out plasma contributions from satellites. As an example, massive companions on non-circular orbits may lead to higher occurrence rates of  tidally-driven heating in close-in satellites orbiting the radio-bright components of binaries in our sample \citep[e.g.][]{Mardling2007MNRAS.382.1768M}. Alternatively,  other mechanisms may on average outweigh the contribution of volcanic activity from satellites. For instance, if binarity enhances the rotation rates of ultracool dwarfs, this could in turn increase the occurrence rate of luminous radiation belts by both enriching ultracool dwarf magnetospheric plasma environments as well as supplying a larger reservoir of energy to power radiation belt electron acceleration mechanisms.

For the first case, increased rotation-driven flaring on ultracool dwarfs may provide a significant source of plasma for their radiation belts. Though \citet{Kao2023Natur.619..272K} showed that quiescent radio emissions are unlikely to directly trace flare emissions, they also  proposed that flares may seed the magnetospheres of the warmest ultracool dwarfs with electrons that then undergo several acceleration stages to reach the MeV energies found in their radiation belts. This hypothesis was motivated by recent studies demonstrating that flares can occur on objects that are at least as late as L5 \citep{Paudel2018ApJ...858...55P, Jackman2019MNRAS.485L.136J, Paudel2020MNRAS.494.5751P, Medina2020ApJ...905..107M}. Ultracool dwarfs appear to flare less frequently and with less energy as they decrease in spectral type \citep{Paudel2018ApJ...858...55P, Medina2020ApJ...905..107M}, so flare contributions to  ultracool dwarf magnetopsheric plasma populations may only be significant for the warmest objects.  In this scenario, ultracool dwarf binaries may experience rotation-enhanced flaring similar to the increased magnetic activity from binary stars attributed to enhanced rotation rates from tidal spin-up \citep{Zahn1977A&A....57..383Z, Morgan2016AJ....151..114M}. Similar phenomena occurring for binary ultracool dwarfs would enrich their magnetospheric plasma environments compared to isolated objects. How rotation drives increased ultracool dwarf flare activity is an open question given key differences in atmospheric ionization and internal structure relative to more massive stars. Nevertheless, rotation increases flare activity even in stars with fully convective dynamo regions \citep{Medina2020ApJ...905..107M} similar to those possessed by ultracool dwarfs.

For the second case, radiation belt activity itself is likely also enhanced by increased rotation rates when all else is equal.  For Jupiter's radiation belts, rotation contributes to various stages of electron acceleration \citep{krupp2007jupi.book..617K, cowleyBunce2001P&SS...49.1067C, kollmann2018JGRA..123.9110K}.  Modeling suggests that rotation is an important energy reservoir for extrasolar radiation belt emissions \citep{leto2021MNRAS.507.1979L, Owocki2022MNRAS.513.1449O} and ultracool dwarf radio detection rates indeed increase with increasing projected rotational velocity $v \sin i$ \citep{Pineda2017ApJ...846...75P}. In fact, isolated T/Y dwarfs may show tentative evidence of a higher quiescent radio occurrence rate compared to L dwarfs, which  \citet{kao2024MNRAS.527.6835K} suggested could be consistent with age-related spin-up \citep{Scholz2018ApJ...859..153S} contributing to the acceleration of electrons in ultracool dwarf radiation belts. In this picture, binarity-related enhancements in the rotation rates of ultracool dwarfs could also contribute to the acceleration of radiation belt electrons and therefore the production of luminous radiation belts.

Finally, we must acknowledge that one possible contributor  to the high binary radio occurrence rate may be that some of the included binary systems are in fact unresolved triples.  With higher multiplicity, a greater fraction of combinations of radio-bright individual components can result in an unresolved system appearing radio-bright as a whole.  However, ultracool dwarf triple systems are very rare  \citep{BardalezGagliuffi2014ApJ...794..143B}, and we have excluded all targets that have evidence of higher-order multiplicity. Our results imply that binaries are excellent targets for radio studies that aim to increase the number of known radio-bright ultracool dwarfs because they may have an elevated radio occurrence rate compared to single objects. 

Gaining further insight into flare contributions to the plasma reservoir traced by ultracool dwarf quiescent radio emissions  will require studies that (1) examine how rotation rates correlate with ultracool dwarf flare rates and (2) compare the quiescent radio occurrence rate for flaring versus non-flaring ultracool dwarfs.  If such studies find a higher occurrence rate of quiescent radio emissions in flaring versus non-flaring objects, this may imply that flares provide a significant source of plasma in the magnetospheres of ultracool dwarfs.

\section{Conclusions}\label{sec.Conclusion}
We present the first detailed statistical comparison of non-flaring quiescent radio emissions in ultracool dwarf binary systems compared to their single-object counterparts.  Prior to this work, detection rate studies did not account for possible effects of binarity.  

After compiling all published radio observations of binary ultracool dwarfs and compare the radio luminosities of binary systems to single objects, we show that although binarity previously appeared to possibly enhance the quiescent radio luminosities of detected ultracool dwarf binary systems relative to single objects, controlling for spectral types may account for this observed phenomenon.

Finally, we show how to apply a generalized analytical Bayesian framework for calculating the occurrence rate of steady emissions in astrophysical populations introduced by \citet{kao2024MNRAS.527.6835K} to comparisons between binary versus single-object systems in a rigorous and self-consistent manner. We then compare the quiescent radio occurrence rate of ultracool dwarf binary systems to single objects with a similar spectral type distribution as for our binary sample and find that binarity enhances the the quiescent radio occurrence rate in ultracool dwarfs compared to their single counterparts.  This enhancement effect is consistent with existing evidence that ultracool dwarf quiescent emissions may largely trace radiation belt emissions \citep{Kao2023Natur.619..272K, Climent2023Sci...381.1120C}.

\section*{Acknowledgements}

MK specially thanks C. Voloshin for valuable statistics and software engineering discussions that shaped this work and A. Schneider for serving as a valuable resource on brown dwarf binaries. She also thanks her Magnetism \& Equity group for their support during the Covid-19 pandemic, and E. Shkolnik for her steadfast mentoring during this difficult time.  Support was provided by NASA through the NASA Hubble Fellowship grant HST-HF2-51411.001-A awarded by the Space Telescope Science Institute, which is operated by the Association of Universities for Research in Astronomy, Inc., for NASA, under contract NAS5-26555; and by the Heising-Simons Foundation through the 51 Pegasi b Fellowship grant 2021-2943. This work is based on observations made with the NSF's Karl G. Jansky Very Large Array (VLA).  This research has made use of the SIMBAD and VizieR databases, operated at CDS, Strasbourg, France; and the European Space Agency (ESA) mission \textit{Gaia} (\url{https://www. cosmos.esa.int/gaia}), processed by the Gaia Data Processing and Analysis Consortium (DPAC, \url{https://www. cosmos.esa.int/web/gaia/dpac/consortium}). This work made use of Astropy (\url{http://www.astropy.org}), a community-developed core Python package and an ecosystem of tools and resources for astronomy \citep{astropy:2013, astropy:2018, astropy:2022};\, Matplotlib \citep{matplotlib};\, Numpy \,\citep{numpy2};\, Scipy \citep{scipy};\, and Scikit-learn \citep{scikit-learn}.


\section*{Data Availability}
This work is a meta-analysis of previously published radio observations of ultracool dwarfs, and all data used in this study have been compiled in table format in \citet{kao2024MNRAS.527.6835K}, \citet{kao2022ApJ...932...21K}, and within this manuscript.
 

\bibliographystyle{mnras}
\bibliography{main} 

\begin{thebibliography}{}
\makeatletter
\relax
\def\mn@urlcharsother{\let\do\@makeother \do\$\do\&\do\#\do\^\do\_\do\%\do\~}
\def\mn@doi{\begingroup\mn@urlcharsother \@ifnextchar [ {\mn@doi@}
  {\mn@doi@[]}}
\def\mn@doi@[#1]#2{\def\@tempa{#1}\ifx\@tempa\@empty \href
  {http://dx.doi.org/#2} {doi:#2}\else \href {http://dx.doi.org/#2} {#1}\fi
  \endgroup}
\def\mn@eprint#1#2{\mn@eprint@#1:#2::\@nil}
\def\mn@eprint@arXiv#1{\href {http://arxiv.org/abs/#1} {{\tt arXiv:#1}}}
\def\mn@eprint@dblp#1{\href {http://dblp.uni-trier.de/rec/bibtex/#1.xml}
  {dblp:#1}}
\def\mn@eprint@#1:#2:#3:#4\@nil{\def\@tempa {#1}\def\@tempb {#2}\def\@tempc
  {#3}\ifx \@tempc \@empty \let \@tempc \@tempb \let \@tempb \@tempa \fi \ifx
  \@tempb \@empty \def\@tempb {arXiv}\fi \@ifundefined
  {mn@eprint@\@tempb}{\@tempb:\@tempc}{\expandafter \expandafter \csname
  mn@eprint@\@tempb\endcsname \expandafter{\@tempc}}}

\bibitem[\protect\citeauthoryear{{Antonova}, {Doyle}, {Hallinan}, {Golden}  \&
  {Koen}}{{Antonova} et~al.}{2007}]{Antonova2007AA...472..257A}
{Antonova} A.,  {Doyle} J.~G.,  {Hallinan} G.,  {Golden} A.,   {Koen} C.,
  2007, \mn@doi [\aap] {10.1051/0004-6361:20077231}, \href
  {https://ui.adsabs.harvard.edu/abs/2007A&A...472..257A} {472, 257}

\bibitem[\protect\citeauthoryear{{Antonova}, {Hallinan}, {Doyle}, {Yu},
  {Kuznetsov}, {Metodieva}, {Golden}  \& {Cruz}}{{Antonova}
  et~al.}{2013}]{Antonova2013AA...549A.131A}
{Antonova} A.,  {Hallinan} G.,  {Doyle} J.~G.,  {Yu} S.,  {Kuznetsov} A.,
  {Metodieva} Y.,  {Golden} A.,   {Cruz} K.~L.,  2013, \mn@doi [\aap]
  {10.1051/0004-6361/201118583}, \href
  {https://ui.adsabs.harvard.edu/abs/2013A&A...549A.131A} {549, A131}

\bibitem[\protect\citeauthoryear{{Artymowicz} \& {Lubow}}{{Artymowicz} \&
  {Lubow}}{1994}]{Artymowicz1994ApJ...421..651A}
{Artymowicz} P.,  {Lubow} S.~H.,  1994, \mn@doi [\apj] {10.1086/173679}, \href
  {https://ui.adsabs.harvard.edu/abs/1994ApJ...421..651A} {421, 651}

\bibitem[\protect\citeauthoryear{{Astropy Collaboration} et~al.,}{{Astropy
  Collaboration} et~al.}{2013}]{astropy:2013}
{Astropy Collaboration} et~al., 2013, \mn@doi [\aap]
  {10.1051/0004-6361/201322068}, \href
  {http://adsabs.harvard.edu/abs/2013A%26A...558A..33A} {558, A33}

\bibitem[\protect\citeauthoryear{{Astropy Collaboration} et~al.,}{{Astropy
  Collaboration} et~al.}{2018}]{astropy:2018}
{Astropy Collaboration} et~al., 2018, \mn@doi [\aj] {10.3847/1538-3881/aabc4f},
  \href {https://ui.adsabs.harvard.edu/abs/2018AJ....156..123A} {156, 123}

\bibitem[\protect\citeauthoryear{{Astropy Collaboration} et~al.,}{{Astropy
  Collaboration} et~al.}{2022}]{astropy:2022}
{Astropy Collaboration} et~al., 2022, \mn@doi [\apj]
  {10.3847/1538-4357/ac7c74}, \href
  {https://ui.adsabs.harvard.edu/abs/2022ApJ...935..167A} {935, 167}

\bibitem[\protect\citeauthoryear{{Audard}, {Brown}, {Briggs}, {G{\"u}del},
  {Telleschi}  \& {Gizis}}{{Audard} et~al.}{2005}]{Audard2005ApJ...625L..63A}
{Audard} M.,  {Brown} A.,  {Briggs} K.~R.,  {G{\"u}del} M.,  {Telleschi} A.,
  {Gizis} J.~E.,  2005, \mn@doi [\apjl] {10.1086/430881}, \href
  {https://ui.adsabs.harvard.edu/abs/2005ApJ...625L..63A} {625, L63}

\bibitem[\protect\citeauthoryear{{Bardalez Gagliuffi} et~al.,}{{Bardalez
  Gagliuffi} et~al.}{2014}]{BardalezGagliuffi2014ApJ...794..143B}
{Bardalez Gagliuffi} D.~C.,  et~al., 2014, \mn@doi [\apj]
  {10.1088/0004-637X/794/2/143}, \href
  {https://ui.adsabs.harvard.edu/abs/2014ApJ...794..143B} {794, 143}

\bibitem[\protect\citeauthoryear{{Bardalez Gagliuffi}, {Gelino}  \&
  {Burgasser}}{{Bardalez Gagliuffi}
  et~al.}{2015}]{BardalezGagliuffi2015AJ....150..163B}
{Bardalez Gagliuffi} D.~C.,  {Gelino} C.~R.,   {Burgasser} A.~J.,  2015,
  \mn@doi [\aj] {10.1088/0004-6256/150/5/163}, \href
  {https://ui.adsabs.harvard.edu/abs/2015AJ....150..163B} {150, 163}

\bibitem[\protect\citeauthoryear{{Basharinov}, {Gurvich}  \&
  {Egorov}}{{Basharinov} et~al.}{1974}]{basharinov1974}
{Basharinov} A.~E.,  {Gurvich} A.~S.,   {Egorov} S.~T.,  1974, Moscow Izdatel
  Nauka, \href {http://adsabs.harvard.edu/abs/1974MoIzN....T....B} {}

\bibitem[\protect\citeauthoryear{{Berger}}{{Berger}}{2002}]{Berger2002ApJ...572..503B}
{Berger} E.,  2002, \mn@doi [\apj] {10.1086/340301}, \href
  {http://adsabs.harvard.edu/abs/2002ApJ...572..503B} {572, 503}

\bibitem[\protect\citeauthoryear{{Berger}}{{Berger}}{2006}]{Berger2006ApJ...648..629B}
{Berger} E.,  2006, \mn@doi [\apj] {10.1086/505787}, \href
  {http://adsabs.harvard.edu/abs/2006ApJ...648..629B} {648, 629}

\bibitem[\protect\citeauthoryear{{Berger} et~al.,}{{Berger}
  et~al.}{2001}]{Berger2001Natur.410..338B}
{Berger} E.,  et~al., 2001, \nat, \href
  {http://adsabs.harvard.edu/abs/2001Natur.410..338B} {410, 338}

\bibitem[\protect\citeauthoryear{{Berger} et~al.,}{{Berger}
  et~al.}{2005}]{Berger2005ApJ...627..960B}
{Berger} E.,  et~al., 2005, \mn@doi [\apj] {10.1086/430343}, \href
  {http://adsabs.harvard.edu/abs/2005ApJ...627..960B} {627, 960}

\bibitem[\protect\citeauthoryear{{Blunt} et~al.,}{{Blunt}
  et~al.}{2017}]{Blunt2017AJ....153..229B}
{Blunt} S.,  et~al., 2017, \mn@doi [\aj] {10.3847/1538-3881/aa6930}, \href
  {https://ui.adsabs.harvard.edu/abs/2017AJ....153..229B} {153, 229}

\bibitem[\protect\citeauthoryear{{Bolton}, {Thorne}, {Bourdarie}, {de Pater}
  \& {Mauk}}{{Bolton} et~al.}{2004}]{bolton2004}
{Bolton} S.~J.,  {Thorne} R.~M.,  {Bourdarie} S.,  {de Pater} I.,   {Mauk} B.,
  2004, {Jupiter's inner radiation belts}.
Cambridge University Press, pp 671--688

\bibitem[\protect\citeauthoryear{{Bouy}, {Brandner}, {Mart{\'\i}n}, {Delfosse},
  {Allard}  \& {Basri}}{{Bouy} et~al.}{2003}]{Bouy2003AJ....126.1526B}
{Bouy} H.,  {Brandner} W.,  {Mart{\'\i}n} E.~L.,  {Delfosse} X.,  {Allard} F.,
   {Basri} G.,  2003, \mn@doi [\aj] {10.1086/377343}, \href
  {https://ui.adsabs.harvard.edu/abs/2003AJ....126.1526B} {126, 1526}

\bibitem[\protect\citeauthoryear{{Bouy} et~al.,}{{Bouy}
  et~al.}{2004}]{Bouy2004AA...423..341B}
{Bouy} H.,  et~al., 2004, \mn@doi [\aap] {10.1051/0004-6361:20040551}, \href
  {https://ui.adsabs.harvard.edu/abs/2004A&A...423..341B} {423, 341}

\bibitem[\protect\citeauthoryear{{Bouy}, {Mart{\'\i}n}, {Brandner}  \&
  {Bouvier}}{{Bouy} et~al.}{2005}]{Bouy2005AJ....129..511B}
{Bouy} H.,  {Mart{\'\i}n} E.~L.,  {Brandner} W.,   {Bouvier} J.,  2005, \mn@doi
  [\aj] {10.1086/426559}, \href
  {https://ui.adsabs.harvard.edu/abs/2005AJ....129..511B} {129, 511}

\bibitem[\protect\citeauthoryear{{Braun}, {Bonaldi}, {Bourke}, {Keane}  \&
  {Wagg}}{{Braun} et~al.}{2019}]{Braun2019arXiv191212699B}
{Braun} R.,  {Bonaldi} A.,  {Bourke} T.,  {Keane} E.,   {Wagg} J.,  2019,
  \mn@doi [arXiv e-prints] {10.48550/arXiv.1912.12699}, \href
  {https://ui.adsabs.harvard.edu/abs/2019arXiv191212699B} {p. arXiv:1912.12699}

\bibitem[\protect\citeauthoryear{{Burgasser} \& {McElwain}}{{Burgasser} \&
  {McElwain}}{2006}]{Burgasser2006AJ....131.1007B}
{Burgasser} A.~J.,  {McElwain} M.~W.,  2006, \mn@doi [\aj] {10.1086/499042},
  \href {https://ui.adsabs.harvard.edu/abs/2006AJ....131.1007B} {131, 1007}

\bibitem[\protect\citeauthoryear{{Burgasser}, {Kirkpatrick}  \&
  {Lowrance}}{{Burgasser} et~al.}{2005}]{Burgasser2005AJ....129.2849B}
{Burgasser} A.~J.,  {Kirkpatrick} J.~D.,   {Lowrance} P.~J.,  2005, \mn@doi
  [\aj] {10.1086/430218}, \href
  {https://ui.adsabs.harvard.edu/abs/2005AJ....129.2849B} {129, 2849}

\bibitem[\protect\citeauthoryear{{Burgasser}, {Cruz}, {Cushing}, {Gelino},
  {Looper}, {Faherty}, {Kirkpatrick}  \& {Reid}}{{Burgasser}
  et~al.}{2010}]{Burgasser2010ApJ...710.1142B}
{Burgasser} A.~J.,  {Cruz} K.~L.,  {Cushing} M.,  {Gelino} C.~R.,  {Looper}
  D.~L.,  {Faherty} J.~K.,  {Kirkpatrick} J.~D.,   {Reid} I.~N.,  2010, \mn@doi
  [\apj] {10.1088/0004-637X/710/2/1142}, \href
  {https://ui.adsabs.harvard.edu/abs/2010ApJ...710.1142B} {710, 1142}

\bibitem[\protect\citeauthoryear{{Burgasser} et~al.,}{{Burgasser}
  et~al.}{2011a}]{Burgasser2011ApJ...735..116B}
{Burgasser} A.~J.,  et~al., 2011a, \mn@doi [\apj]
  {10.1088/0004-637X/735/2/116}, \href
  {https://ui.adsabs.harvard.edu/abs/2011ApJ...735..116B} {735, 116}

\bibitem[\protect\citeauthoryear{{Burgasser}, {Sitarski}, {Gelino}, {Logsdon}
  \& {Perrin}}{{Burgasser} et~al.}{2011b}]{Burgasser2011ApJ...739...49B}
{Burgasser} A.~J.,  {Sitarski} B.~N.,  {Gelino} C.~R.,  {Logsdon} S.~E.,
  {Perrin} M.~D.,  2011b, \mn@doi [\apj] {10.1088/0004-637X/739/1/49}, \href
  {https://ui.adsabs.harvard.edu/abs/2011ApJ...739...49B} {739, 49}

\bibitem[\protect\citeauthoryear{{Burgasser}, {Melis}, {Zauderer}  \&
  {Berger}}{{Burgasser} et~al.}{2013a}]{Burgasser2013ApJ...762L...3B}
{Burgasser} A.~J.,  {Melis} C.,  {Zauderer} B.~A.,   {Berger} E.,  2013a,
  \mn@doi [\apjl] {10.1088/2041-8205/762/1/L3}, \href
  {http://adsabs.harvard.edu/abs/2013ApJ...762L...3B} {762, L3}

\bibitem[\protect\citeauthoryear{{Burgasser}, {Sheppard}  \&
  {Luhman}}{{Burgasser} et~al.}{2013b}]{Burgasser2013ApJ...772..129B}
{Burgasser} A.~J.,  {Sheppard} S.~S.,   {Luhman} K.~L.,  2013b, \mn@doi [\apj]
  {10.1088/0004-637X/772/2/129}, \href
  {https://ui.adsabs.harvard.edu/abs/2013ApJ...772..129B} {772, 129}

\bibitem[\protect\citeauthoryear{{Burgasser} et~al.,}{{Burgasser}
  et~al.}{2015a}]{Burgasser2015AJ....149..104B}
{Burgasser} A.~J.,  et~al., 2015a, \mn@doi [\aj] {10.1088/0004-6256/149/3/104},
  \href {http://cdsads.u-strasbg.fr/abs/2015AJ....149..104B} {149, 104}

\bibitem[\protect\citeauthoryear{{Burgasser}, {Melis}, {Todd}, {Gelino},
  {Hallinan}  \& {Bardalez Gagliuffi}}{{Burgasser}
  et~al.}{2015b}]{Burgasser2015AJ....150..180B}
{Burgasser} A.~J.,  {Melis} C.,  {Todd} J.,  {Gelino} C.~R.,  {Hallinan} G.,
  {Bardalez Gagliuffi} D.,  2015b, \mn@doi [\aj] {10.1088/0004-6256/150/6/180},
  \href {https://ui.adsabs.harvard.edu/abs/2015AJ....150..180B} {150, 180}

\bibitem[\protect\citeauthoryear{{Callingham} et~al.,}{{Callingham}
  et~al.}{2021}]{Callingham2021NatAs...5.1233C}
{Callingham} J.~R.,  et~al., 2021, \mn@doi [Nature Astronomy]
  {10.1038/s41550-021-01483-0}, \href
  {https://ui.adsabs.harvard.edu/abs/2021NatAs...5.1233C} {5, 1233}

\bibitem[\protect\citeauthoryear{{Canup} \& {Ward}}{{Canup} \&
  {Ward}}{2006}]{CanupWard2006Natur.441..834C}
{Canup} R.~M.,  {Ward} W.~R.,  2006, \mn@doi [\nat] {10.1038/nature04860},
  \href {https://ui.adsabs.harvard.edu/abs/2006Natur.441..834C} {441, 834}

\bibitem[\protect\citeauthoryear{{Cilibrasi}, {Szul{\'a}gyi}, {Grimm}  \&
  {Mayer}}{{Cilibrasi} et~al.}{2021}]{Cilibrasi2021MNRAS.504.5455C}
{Cilibrasi} M.,  {Szul{\'a}gyi} J.,  {Grimm} S.~L.,   {Mayer} L.,  2021,
  \mn@doi [\mnras] {10.1093/mnras/stab1179}, \href
  {https://ui.adsabs.harvard.edu/abs/2021MNRAS.504.5455C} {504, 5455}

\bibitem[\protect\citeauthoryear{{Clarke}, {Grodent}, {Cowley}, {Bunce},
  {Zarka}, {Connerney}  \& {Satoh}}{{Clarke}
  et~al.}{2004}]{Clarke2004jpsm.book..639C}
{Clarke} J.~T.,  {Grodent} D.,  {Cowley} S.~W.~H.,  {Bunce} E.~J.,  {Zarka} P.,
   {Connerney} J.~E.~P.,   {Satoh} T.,  2004, {Jupiter's aurora}.
Cambridge University Press, pp 639--670

\bibitem[\protect\citeauthoryear{{Climent} et~al.,}{{Climent}
  et~al.}{2022}]{climent2022AA...660A..65C}
{Climent} J.~B.,  et~al., 2022, \mn@doi [\aap] {10.1051/0004-6361/202142260},
  \href {https://ui.adsabs.harvard.edu/abs/2022A&A...660A..65C} {660, A65}

\bibitem[\protect\citeauthoryear{{Climent}, {Guirado}, {P{\'e}rez-Torres},
  {Marcaide}  \& {Pe{\~n}a-Mo{\~n}ino}}{{Climent}
  et~al.}{2023}]{Climent2023Sci...381.1120C}
{Climent} J.~B.,  {Guirado} J.~C.,  {P{\'e}rez-Torres} M.,  {Marcaide} J.~M.,
  {Pe{\~n}a-Mo{\~n}ino} L.,  2023, \mn@doi [Science] {10.1126/science.adg6635},
  \href {https://ui.adsabs.harvard.edu/abs/2023Sci...381.1120C} {381, 1120}

\bibitem[\protect\citeauthoryear{{Cowley} \& {Bunce}}{{Cowley} \&
  {Bunce}}{2001}]{cowleyBunce2001P&SS...49.1067C}
{Cowley} S.~W.~H.,  {Bunce} E.~J.,  2001, \mn@doi [\planss]
  {10.1016/S0032-0633(00)00167-7}, \href
  {https://ui.adsabs.harvard.edu/abs/2001P&SS...49.1067C} {49, 1067}

\bibitem[\protect\citeauthoryear{{Deshpande} et~al.,}{{Deshpande}
  et~al.}{2012}]{Deshpande2012AJ....144...99D}
{Deshpande} R.,  et~al., 2012, \mn@doi [\aj] {10.1088/0004-6256/144/4/99},
  \href {https://ui.adsabs.harvard.edu/abs/2012AJ....144...99D} {144, 99}

\bibitem[\protect\citeauthoryear{{Dieterich}, {Henry}, {Jao}, {Winters},
  {Hosey}, {Riedel}  \& {Subasavage}}{{Dieterich}
  et~al.}{2014}]{Dieterich2014AJ....147...94D}
{Dieterich} S.~B.,  {Henry} T.~J.,  {Jao} W.-C.,  {Winters} J.~G.,  {Hosey}
  A.~D.,  {Riedel} A.~R.,   {Subasavage} J.~P.,  2014, \mn@doi [\aj]
  {10.1088/0004-6256/147/5/94}, \href
  {https://ui.adsabs.harvard.edu/abs/2014AJ....147...94D} {147, 94}

\bibitem[\protect\citeauthoryear{{Dieterich} et~al.,}{{Dieterich}
  et~al.}{2018}]{Dieterich2018ApJ...865...28D}
{Dieterich} S.~B.,  et~al., 2018, \mn@doi [\apj] {10.3847/1538-4357/aadadc},
  \href {https://ui.adsabs.harvard.edu/abs/2018ApJ...865...28D} {865, 28}

\bibitem[\protect\citeauthoryear{{Dupuy} \& {Liu}}{{Dupuy} \&
  {Liu}}{2012}]{Dupuy2012ApJS..201...19D}
{Dupuy} T.~J.,  {Liu} M.~C.,  2012, \mn@doi [\apjs]
  {10.1088/0067-0049/201/2/19}, \href
  {https://ui.adsabs.harvard.edu/abs/2012ApJS..201...19D} {201, 19}

\bibitem[\protect\citeauthoryear{{Dupuy} \& {Liu}}{{Dupuy} \&
  {Liu}}{2017}]{Dupuy2017ApJS..231...15D}
{Dupuy} T.~J.,  {Liu} M.~C.,  2017, \mn@doi [\apjs] {10.3847/1538-4365/aa5e4c},
  \href {https://ui.adsabs.harvard.edu/abs/2017ApJS..231...15D} {231, 15}

\bibitem[\protect\citeauthoryear{{Dupuy}, {Liu}  \& {Ireland}}{{Dupuy}
  et~al.}{2009}]{Dupuy2009ApJ...692..729D}
{Dupuy} T.~J.,  {Liu} M.~C.,   {Ireland} M.~J.,  2009, \mn@doi [\apj]
  {10.1088/0004-637X/692/1/729}, \href
  {https://ui.adsabs.harvard.edu/abs/2009ApJ...692..729D} {692, 729}

\bibitem[\protect\citeauthoryear{{Dupuy} et~al.,}{{Dupuy}
  et~al.}{2019}]{Dupuy2019AJ....158..174D}
{Dupuy} T.~J.,  et~al., 2019, \mn@doi [\aj] {10.3847/1538-3881/ab3cd1}, \href
  {https://ui.adsabs.harvard.edu/abs/2019AJ....158..174D} {158, 174}

\bibitem[\protect\citeauthoryear{{Faherty}, {Burgasser}, {Cruz}, {Shara},
  {Walter}  \& {Gelino}}{{Faherty} et~al.}{2009}]{Faherty2009AJ....137....1F}
{Faherty} J.~K.,  {Burgasser} A.~J.,  {Cruz} K.~L.,  {Shara} M.~M.,  {Walter}
  F.~M.,   {Gelino} C.~R.,  2009, \mn@doi [\aj] {10.1088/0004-6256/137/1/1},
  \href {https://ui.adsabs.harvard.edu/abs/2009AJ....137....1F} {137, 1}

\bibitem[\protect\citeauthoryear{{Faherty} et~al.,}{{Faherty}
  et~al.}{2012}]{Faherty2012ApJ...752...56F}
{Faherty} J.~K.,  et~al., 2012, \mn@doi [\apj] {10.1088/0004-637X/752/1/56},
  \href {https://ui.adsabs.harvard.edu/abs/2012ApJ...752...56F} {752, 56}

\bibitem[\protect\citeauthoryear{{Feigelson} \& Jogesh}{{Feigelson} \&
  Jogesh}{2012}]{FeigelsonAstroStats}
{Feigelson} E.~D.,  Jogesh B.~G.,  2012, Modern Statistical Methods for
  Astronomy: With R Applications, 1st ed. edn.
Cambridge University Press

\bibitem[\protect\citeauthoryear{{Forbrich} et~al.,}{{Forbrich}
  et~al.}{2016}]{Forbrich2016ApJ...827...22F}
{Forbrich} J.,  et~al., 2016, \mn@doi [\apj] {10.3847/0004-637X/827/1/22},
  \href {https://ui.adsabs.harvard.edu/abs/2016ApJ...827...22F} {827, 22}

\bibitem[\protect\citeauthoryear{{Forrest}, {Skrutskie}  \& {Shure}}{{Forrest}
  et~al.}{1988}]{Forrest1988ApJ...330L.119F}
{Forrest} W.~J.,  {Skrutskie} M.~F.,   {Shure} M.,  1988, \mn@doi [\apjl]
  {10.1086/185218}, \href
  {https://ui.adsabs.harvard.edu/abs/1988ApJ...330L.119F} {330, L119}

\bibitem[\protect\citeauthoryear{{Gaia Collaboration} et~al.,}{{Gaia
  Collaboration} et~al.}{2018}]{Gaia20182018AA...616A...1G}
{Gaia Collaboration} et~al., 2018, \mn@doi [\aap]
  {10.1051/0004-6361/201833051}, \href
  {https://ui.adsabs.harvard.edu/abs/2018A&A...616A...1G} {616, A1}

\bibitem[\protect\citeauthoryear{{Ganushkina}, {Dandouras}, {Shprits}  \&
  {Cao}}{{Ganushkina} et~al.}{2011}]{ganushkina2011}
{Ganushkina} N.~Y.,  {Dandouras} I.,  {Shprits} Y.~Y.,   {Cao} J.,  2011,
  \mn@doi [Journal of Geophysical Research (Space Physics)]
  {10.1029/2010JA016376}, \href
  {http://adsabs.harvard.edu/abs/2011JGRA..116.9234G} {116, A09234}

\bibitem[\protect\citeauthoryear{{Gauza}, {B{\'e}jar}, {P{\'e}rez-Garrido},
  {Zapatero Osorio}, {Lodieu}, {Rebolo}, {Pall{\'e}}  \& {Nowak}}{{Gauza}
  et~al.}{2015}]{Gauza2015ApJ...804...96G}
{Gauza} B.,  {B{\'e}jar} V. J.~S.,  {P{\'e}rez-Garrido} A.,  {Zapatero Osorio}
  M.~R.,  {Lodieu} N.,  {Rebolo} R.,  {Pall{\'e}} E.,   {Nowak} G.,  2015,
  \mn@doi [\apj] {10.1088/0004-637X/804/2/96}, \href
  {https://ui.adsabs.harvard.edu/abs/2015ApJ...804...96G} {804, 96}

\bibitem[\protect\citeauthoryear{{Gelino} et~al.,}{{Gelino}
  et~al.}{2014}]{Gelino2014AJ....148....6G}
{Gelino} C.~R.,  et~al., 2014, \mn@doi [\aj] {10.1088/0004-6256/148/1/6}, \href
  {https://ui.adsabs.harvard.edu/abs/2014AJ....148....6G} {148, 6}

\bibitem[\protect\citeauthoryear{{Gillon} et~al.,}{{Gillon}
  et~al.}{2017}]{Gillon2017Natur.542..456G}
{Gillon} M.,  et~al., 2017, \mn@doi [\nat] {10.1038/nature21360}, \href
  {https://ui.adsabs.harvard.edu/abs/2017Natur.542..456G} {542, 456}

\bibitem[\protect\citeauthoryear{{Girard} et~al.,}{{Girard}
  et~al.}{2016}]{girard2016}
{Girard} J.~N.,  et~al., 2016, \mn@doi [\aap] {10.1051/0004-6361/201527518},
  \href {http://adsabs.harvard.edu/abs/2016A%26A...587A...3G} {587, A3}

\bibitem[\protect\citeauthoryear{{Gizis}}{{Gizis}}{2002}]{Gizis2002ApJ...575..484G}
{Gizis} J.~E.,  2002, \mn@doi [\apj] {10.1086/341259}, \href
  {https://ui.adsabs.harvard.edu/abs/2002ApJ...575..484G} {575, 484}

\bibitem[\protect\citeauthoryear{{Gizis}, {Burgasser}, {Berger}, {Williams},
  {Vrba}, {Cruz}  \& {Metchev}}{{Gizis}
  et~al.}{2013}]{Gizis2013ApJ...779..172G}
{Gizis} J.~E.,  {Burgasser} A.~J.,  {Berger} E.,  {Williams} P. K.~G.,  {Vrba}
  F.~J.,  {Cruz} K.~L.,   {Metchev} S.,  2013, \mn@doi [\apj]
  {10.1088/0004-637X/779/2/172}, \href
  {https://ui.adsabs.harvard.edu/abs/2013ApJ...779..172G} {779, 172}

\bibitem[\protect\citeauthoryear{{Golimowski} et~al.,}{{Golimowski}
  et~al.}{2004}]{Golimowski2004AJ....128.1733G}
{Golimowski} D.~A.,  et~al., 2004, \mn@doi [\aj] {10.1086/423911}, \href
  {https://ui.adsabs.harvard.edu/abs/2004AJ....128.1733G} {128, 1733}

\bibitem[\protect\citeauthoryear{{Guedel} \& {Benz}}{{Guedel} \&
  {Benz}}{1993}]{Guedel1993ApJ...405L..63G}
{Guedel} M.,  {Benz} A.~O.,  1993, \mn@doi [\apjl] {10.1086/186766}, \href
  {https://ui.adsabs.harvard.edu/abs/1993ApJ...405L..63G} {405, L63}

\bibitem[\protect\citeauthoryear{{Guenther} \& {Wuchterl}}{{Guenther} \&
  {Wuchterl}}{2003}]{Guenther2003AA...401..677G}
{Guenther} E.~W.,  {Wuchterl} G.,  2003, \mn@doi [\aap]
  {10.1051/0004-6361:20030149}, \href
  {https://ui.adsabs.harvard.edu/abs/2003A&A...401..677G} {401, 677}

\bibitem[\protect\citeauthoryear{{Guirado}, {Azulay}, {Gauza},
  {P{\'e}rez-Torres}, {Rebolo}, {Climent}  \& {Zapatero Osorio}}{{Guirado}
  et~al.}{2018}]{Guirado2018AA...610A..23G}
{Guirado} J.~C.,  {Azulay} R.,  {Gauza} B.,  {P{\'e}rez-Torres} M.~A.,
  {Rebolo} R.,  {Climent} J.~B.,   {Zapatero Osorio} M.~R.,  2018, \mn@doi
  [\aap] {10.1051/0004-6361/201732130}, \href
  {https://ui.adsabs.harvard.edu/abs/2018A&A...610A..23G} {610, A23}

\bibitem[\protect\citeauthoryear{{Haghighipour} \& {Raymond}}{{Haghighipour} \&
  {Raymond}}{2007}]{Haghighipour2007ApJ...666..436H}
{Haghighipour} N.,  {Raymond} S.~N.,  2007, \mn@doi [\apj] {10.1086/520501},
  \href {https://ui.adsabs.harvard.edu/abs/2007ApJ...666..436H} {666, 436}

\bibitem[\protect\citeauthoryear{{Hallinan}, {Antonova}, {Doyle}, {Bourke},
  {Brisken}  \& {Golden}}{{Hallinan}
  et~al.}{2006}]{Hallinan2006ApJ...653..690H}
{Hallinan} G.,  {Antonova} A.,  {Doyle} J.~G.,  {Bourke} S.,  {Brisken} W.~F.,
   {Golden} A.,  2006, \mn@doi [\apj] {10.1086/508678}, \href
  {http://adsabs.harvard.edu/abs/2006ApJ...653..690H} {653, 690}

\bibitem[\protect\citeauthoryear{{Hallinan} et~al.,}{{Hallinan}
  et~al.}{2007}]{Hallinan2007ApJ...663L..25H}
{Hallinan} G.,  et~al., 2007, \mn@doi [\apjl] {10.1086/519790}, \href
  {http://adsabs.harvard.edu/abs/2007ApJ...663L..25H} {663, L25}

\bibitem[\protect\citeauthoryear{{Hallinan}, {Antonova}, {Doyle}, {Bourke},
  {Lane}  \& {Golden}}{{Hallinan} et~al.}{2008}]{Hallinan2008ApJ...684..644H}
{Hallinan} G.,  {Antonova} A.,  {Doyle} J.~G.,  {Bourke} S.,  {Lane} C.,
  {Golden} A.,  2008, \mn@doi [\apj] {10.1086/590360}, \href
  {http://adsabs.harvard.edu/abs/2008ApJ...684..644H} {684, 644}

\bibitem[\protect\citeauthoryear{{Hallinan} et~al.,}{{Hallinan}
  et~al.}{2015}]{Hallinan2015Natur.523..568H}
{Hallinan} G.,  et~al., 2015, \mn@doi [\nat] {10.1038/nature14619}, \href
  {http://adsabs.harvard.edu/abs/2015Natur.523..568H} {523, 568}

\bibitem[\protect\citeauthoryear{Hallinan et~al.,}{Hallinan
  et~al.}{2019}]{hallinan2019dsa2000}
Hallinan G.,  et~al., 2019, The DSA-2000 -- A Radio Survey Camera (\mn@eprint
  {arXiv} {1907.07648})

\bibitem[\protect\citeauthoryear{{Hardegree-Ullman}, {Cushing}, {Muirhead}  \&
  {Christiansen}}{{Hardegree-Ullman}
  et~al.}{2019}]{hardegree-ullman2019AJ....158...75H}
{Hardegree-Ullman} K.~K.,  {Cushing} M.~C.,  {Muirhead} P.~S.,   {Christiansen}
  J.~L.,  2019, \mn@doi [\aj] {10.3847/1538-3881/ab21d2}, \href
  {https://ui.adsabs.harvard.edu/abs/2019AJ....158...75H} {158, 75}

\bibitem[\protect\citeauthoryear{{He}, {Triaud}  \& {Gillon}}{{He}
  et~al.}{2017}]{He2017MNRAS.464.2687H}
{He} M.~Y.,  {Triaud} A.~H.~M.~J.,   {Gillon} M.,  2017, \mn@doi [\mnras]
  {10.1093/mnras/stw2391}, \href
  {http://adsabs.harvard.edu/abs/2017MNRAS.464.2687H} {464, 2687}

\bibitem[\protect\citeauthoryear{Horne, Thorne, Glauert, Douglas~Menietti,
  Shprits  \& Gurnett}{Horne et~al.}{2008}]{Horne2008}
Horne R.~B.,  Thorne R.~M.,  Glauert S.~A.,  Douglas~Menietti J.,  Shprits
  Y.~Y.,   Gurnett D.~A.,  2008, Nature Physics, 4, 301 EP

\bibitem[\protect\citeauthoryear{{Hughes}, {Boley}, {Osten}, {White}  \&
  {Leacock}}{{Hughes} et~al.}{2021}]{hughes2021AJ....162...43H}
{Hughes} A.~G.,  {Boley} A.~C.,  {Osten} R.~A.,  {White} J.~A.,   {Leacock} M.,
   2021, \mn@doi [\aj] {10.3847/1538-3881/ac02c3}, \href
  {https://ui.adsabs.harvard.edu/abs/2021AJ....162...43H} {162, 43}

\bibitem[\protect\citeauthoryear{{Hunter}}{{Hunter}}{2007}]{matplotlib}
{Hunter} J.~D.,  2007, \mn@doi [Computing in Science and Engineering]
  {10.1109/MCSE.2007.55}, \href
  {http://adsabs.harvard.edu/abs/2007CSE.....9...90H} {9, 90}

\bibitem[\protect\citeauthoryear{{Jackman} et~al.,}{{Jackman}
  et~al.}{2019}]{Jackman2019MNRAS.485L.136J}
{Jackman} J. A.~G.,  et~al., 2019, \mn@doi [\mnras] {10.1093/mnrasl/slz039},
  \href {https://ui.adsabs.harvard.edu/abs/2019MNRAS.485L.136J} {485, L136}

\bibitem[\protect\citeauthoryear{{Kao} \& {Pineda}}{{Kao} \&
  {Pineda}}{2022}]{kao2022ApJ...932...21K}
{Kao} M.~M.,  {Pineda} J.~S.,  2022, \mn@doi [\apj] {10.3847/1538-4357/ac660b},
  \href {https://ui.adsabs.harvard.edu/abs/2022ApJ...932...21K} {932, 21}

\bibitem[\protect\citeauthoryear{{Kao} \& {Shkolnik}}{{Kao} \&
  {Shkolnik}}{2024}]{kao2024MNRAS.527.6835K}
{Kao} M.~M.,  {Shkolnik} E.~L.,  2024, \mn@doi [\mnras]
  {10.1093/mnras/stad2272}, \href
  {https://ui.adsabs.harvard.edu/abs/2024MNRAS.527.6835K} {527, 6835}

\bibitem[\protect\citeauthoryear{{Kao}, {Hallinan}, {Pineda}, {Escala},
  {Burgasser}, {Bourke}  \& {Stevenson}}{{Kao}
  et~al.}{2016}]{Kao2016ApJ...818...24K}
{Kao} M.~M.,  {Hallinan} G.,  {Pineda} J.~S.,  {Escala} I.,  {Burgasser} A.,
  {Bourke} S.,   {Stevenson} D.,  2016, \mn@doi [\apj]
  {10.3847/0004-637X/818/1/24}, \href
  {http://adsabs.harvard.edu/abs/2016ApJ...818...24K} {818, 24}

\bibitem[\protect\citeauthoryear{{Kao}, {Hallinan}, {Pineda}, {Stevenson}  \&
  {Burgasser}}{{Kao} et~al.}{2018}]{Kao2018ApJS..237...25K}
{Kao} M.~M.,  {Hallinan} G.,  {Pineda} J.~S.,  {Stevenson} D.,   {Burgasser}
  A.,  2018, \mn@doi [\apjs] {10.3847/1538-4365/aac2d5}, \href
  {https://ui.adsabs.harvard.edu/abs/2018ApJS..237...25K} {237, 25}

\bibitem[\protect\citeauthoryear{{Kao}, {Hallinan}  \& {Pineda}}{{Kao}
  et~al.}{2019}]{Kao2019MNRAS.487.1994K}
{Kao} M.~M.,  {Hallinan} G.,   {Pineda} J.~S.,  2019, \mn@doi [\mnras]
  {10.1093/mnras/stz1372}, \href
  {https://ui.adsabs.harvard.edu/abs/2019MNRAS.487.1994K} {487, 1994}

\bibitem[\protect\citeauthoryear{{Kao}, {Mioduszewski}, {Villadsen}  \&
  {Shkolnik}}{{Kao} et~al.}{2023}]{Kao2023Natur.619..272K}
{Kao} M.~M.,  {Mioduszewski} A.~J.,  {Villadsen} J.,   {Shkolnik} E.~L.,  2023,
  \mn@doi [\nat] {10.1038/s41586-023-06138-w}, \href
  {https://ui.adsabs.harvard.edu/abs/2023Natur.619..272K} {619, 272}

\bibitem[\protect\citeauthoryear{Kaplan \& Meier}{Kaplan \&
  Meier}{1958}]{Kaplan-Meier1958}
Kaplan E.~L.,  Meier P.,  1958, Journal of the American Statistical
  Association, 53, 457

\bibitem[\protect\citeauthoryear{{Kellermann}}{{Kellermann}}{1970}]{kellerman1970}
{Kellermann} K.~I.,  1970, \mn@doi [Radio Science] {10.1029/RS005i002p00487},
  \href {http://adsabs.harvard.edu/abs/1970RaSc....5..487K} {5, 487}

\bibitem[\protect\citeauthoryear{{Khurana}, {Kivelson}, {Vasyliunas}, {Krupp},
  {Woch}, {Lagg}, {Mauk}  \& {Kurth}}{{Khurana} et~al.}{2004}]{khurana2004}
{Khurana} K.~K.,  {Kivelson} M.~G.,  {Vasyliunas} V.~M.,  {Krupp} N.,  {Woch}
  J.,  {Lagg} A.,  {Mauk} B.~H.,   {Kurth} W.~S.,  2004, {The configuration of
  Jupiter's magnetosphere}.
Cambridge University Press, pp 593--616

\bibitem[\protect\citeauthoryear{{Kirkpatrick} et~al.,}{{Kirkpatrick}
  et~al.}{2000}]{Kirkpatrick2000AJ....120..447K}
{Kirkpatrick} J.~D.,  et~al., 2000, \mn@doi [\aj] {10.1086/301427}, \href
  {https://ui.adsabs.harvard.edu/abs/2000AJ....120..447K} {120, 447}

\bibitem[\protect\citeauthoryear{{Koen}}{{Koen}}{2013}]{Koen2013MNRAS.428.2824K}
{Koen} C.,  2013, \mn@doi [\mnras] {10.1093/mnras/sts208}, \href
  {https://ui.adsabs.harvard.edu/abs/2013MNRAS.428.2824K} {428, 2824}

\bibitem[\protect\citeauthoryear{{K{\"o}hler}, {Ratzka}  \&
  {Leinert}}{{K{\"o}hler} et~al.}{2012}]{Kohler2012AA...541A..29K}
{K{\"o}hler} R.,  {Ratzka} T.,   {Leinert} C.,  2012, \mn@doi [\aap]
  {10.1051/0004-6361/201118707}, \href
  {https://ui.adsabs.harvard.edu/abs/2012A&A...541A..29K} {541, A29}

\bibitem[\protect\citeauthoryear{{Kollmann}, {Roussos}, {Paranicas},
  {Woodfield}, {Mauk}, {Clark}, {Smith}  \& {Vandegriff}}{{Kollmann}
  et~al.}{2018}]{kollmann2018JGRA..123.9110K}
{Kollmann} P.,  {Roussos} E.,  {Paranicas} C.,  {Woodfield} E.~E.,  {Mauk}
  B.~H.,  {Clark} G.,  {Smith} D.~C.,   {Vandegriff} J.,  2018, \mn@doi
  [Journal of Geophysical Research (Space Physics)] {10.1029/2018JA025665},
  \href {https://ui.adsabs.harvard.edu/abs/2018JGRA..123.9110K} {123, 9110}

\bibitem[\protect\citeauthoryear{{Konopacky}, {Ghez}, {Barman}, {Rice},
  {Bailey}, {White}, {McLean}  \& {Duch{\^e}ne}}{{Konopacky}
  et~al.}{2010}]{Konopacky2010ApJ...711.1087K}
{Konopacky} Q.~M.,  {Ghez} A.~M.,  {Barman} T.~S.,  {Rice} E.~L.,  {Bailey}
  J.~I. I.,  {White} R.~J.,  {McLean} I.~S.,   {Duch{\^e}ne} G.,  2010, \mn@doi
  [\apj] {10.1088/0004-637X/711/2/1087}, \href
  {https://ui.adsabs.harvard.edu/abs/2010ApJ...711.1087K} {711, 1087}

\bibitem[\protect\citeauthoryear{{Konopacky} et~al.,}{{Konopacky}
  et~al.}{2012}]{Konopacky2012ApJ...750...79K}
{Konopacky} Q.~M.,  et~al., 2012, \mn@doi [\apj] {10.1088/0004-637X/750/1/79},
  \href {https://ui.adsabs.harvard.edu/abs/2012ApJ...750...79K} {750, 79}

\bibitem[\protect\citeauthoryear{{Krishnamurthi}, {Leto}  \&
  {Linsky}}{{Krishnamurthi} et~al.}{1999}]{Krishnamurthi1999AJ....118.1369K}
{Krishnamurthi} A.,  {Leto} G.,   {Linsky} J.~L.,  1999, \mn@doi [\aj]
  {10.1086/301015}, \href
  {https://ui.adsabs.harvard.edu/abs/1999AJ....118.1369K} {118, 1369}

\bibitem[\protect\citeauthoryear{{Krupp} et~al.,}{{Krupp}
  et~al.}{2007}]{krupp2007jupi.book..617K}
{Krupp} N.,  et~al., 2007, in , Jupiter.
Cambridge University Press, Chapt.~25, pp 617--638

\bibitem[\protect\citeauthoryear{{Lanza}}{{Lanza}}{2012}]{Lanza2012A&A...544A..23L}
{Lanza} A.~F.,  2012, \mn@doi [\aap] {10.1051/0004-6361/201219002}, \href
  {https://ui.adsabs.harvard.edu/abs/2012A&A...544A..23L} {544, A23}

\bibitem[\protect\citeauthoryear{{Lazorenko} \& {Sahlmann}}{{Lazorenko} \&
  {Sahlmann}}{2018}]{Lazorenko2018AA...618A.111L}
{Lazorenko} P.~F.,  {Sahlmann} J.,  2018, \mn@doi [\aap]
  {10.1051/0004-6361/201833626}, \href
  {https://ui.adsabs.harvard.edu/abs/2018A&A...618A.111L} {618, A111}

\bibitem[\protect\citeauthoryear{{Lee} \& {Wang}}{{Lee} \&
  {Wang}}{2003}]{Lee_survival}
{Lee} E.~T.,  {Wang} J.~W.,  2003, Statistical methods for survival data
  analysis, 3rd ed. edn.
Wiley series in probability and statistics, J. Wiley, New York

\bibitem[\protect\citeauthoryear{{Leinert}, {Allard}, {Richichi}  \&
  {Hauschildt}}{{Leinert} et~al.}{2000}]{Leinert2000AA...353..691L}
{Leinert} C.,  {Allard} F.,  {Richichi} A.,   {Hauschildt} P.~H.,  2000, \aap,
  \href {https://ui.adsabs.harvard.edu/abs/2000A&A...353..691L} {353, 691}

\bibitem[\protect\citeauthoryear{{Leto} et~al.,}{{Leto}
  et~al.}{2021}]{leto2021MNRAS.507.1979L}
{Leto} P.,  et~al., 2021, \mn@doi [\mnras] {10.1093/mnras/stab2168}, \href
  {https://ui.adsabs.harvard.edu/abs/2021MNRAS.507.1979L} {507, 1979}

\bibitem[\protect\citeauthoryear{{Liu}, {Dupuy}  \& {Allers}}{{Liu}
  et~al.}{2016}]{Liu2016ApJ...833...96L}
{Liu} M.~C.,  {Dupuy} T.~J.,   {Allers} K.~N.,  2016, \mn@doi [\apj]
  {10.3847/1538-4357/833/1/96}, \href
  {https://ui.adsabs.harvard.edu/abs/2016ApJ...833...96L} {833, 96}

\bibitem[\protect\citeauthoryear{{Luhman}}{{Luhman}}{2013}]{Luhman2013ApJ...767L...1L}
{Luhman} K.~L.,  2013, \mn@doi [\apjl] {10.1088/2041-8205/767/1/L1}, \href
  {https://ui.adsabs.harvard.edu/abs/2013ApJ...767L...1L} {767, L1}

\bibitem[\protect\citeauthoryear{{Lynch}, {Murphy}, {Ravi}, {Hobbs}, {Lo}  \&
  {Ward}}{{Lynch} et~al.}{2016}]{Lynch2016MNRAS.457.1224L}
{Lynch} C.,  {Murphy} T.,  {Ravi} V.,  {Hobbs} G.,  {Lo} K.,   {Ward} C.,
  2016, \mn@doi [\mnras] {10.1093/mnras/stw050}, \href
  {http://adsabs.harvard.edu/abs/2016MNRAS.457.1224L} {457, 1224}

\bibitem[\protect\citeauthoryear{{Mardling}}{{Mardling}}{2007}]{Mardling2007MNRAS.382.1768M}
{Mardling} R.~A.,  2007, \mn@doi [\mnras] {10.1111/j.1365-2966.2007.12500.x},
  \href {https://ui.adsabs.harvard.edu/abs/2007MNRAS.382.1768M} {382, 1768}

\bibitem[\protect\citeauthoryear{{Mauk} \& {Fox}}{{Mauk} \&
  {Fox}}{2010}]{MaukFox2010JGRA..11512220M}
{Mauk} B.~H.,  {Fox} N.~J.,  2010, \mn@doi [Journal of Geophysical Research
  (Space Physics)] {10.1029/2010JA015660}, \href
  {https://ui.adsabs.harvard.edu/abs/2010JGRA..11512220M} {115, A12220}

\bibitem[\protect\citeauthoryear{{McCaughrean}, {Scholz}  \&
  {Lodieu}}{{McCaughrean} et~al.}{2002}]{McCaughrean2002AA...390L..27M}
{McCaughrean} M.~J.,  {Scholz} R.~D.,   {Lodieu} N.,  2002, \mn@doi [\aap]
  {10.1051/0004-6361:20020928}, \href
  {https://ui.adsabs.harvard.edu/abs/2002A&A...390L..27M} {390, L27}

\bibitem[\protect\citeauthoryear{{McElwain} \& {Burgasser}}{{McElwain} \&
  {Burgasser}}{2006}]{McElwain2006AJ....132.2074M}
{McElwain} M.~W.,  {Burgasser} A.~J.,  2006, \mn@doi [\aj] {10.1086/508199},
  \href {https://ui.adsabs.harvard.edu/abs/2006AJ....132.2074M} {132, 2074}

\bibitem[\protect\citeauthoryear{{McLean}, {Berger}, {Irwin}, {Forbrich}  \&
  {Reiners}}{{McLean} et~al.}{2011}]{McLean2011ApJ...741...27M}
{McLean} M.,  {Berger} E.,  {Irwin} J.,  {Forbrich} J.,   {Reiners} A.,  2011,
  \mn@doi [\apj] {10.1088/0004-637X/741/1/27}, \href
  {http://adsabs.harvard.edu/abs/2011ApJ...741...27M} {741, 27}

\bibitem[\protect\citeauthoryear{{McLean}, {Berger}  \& {Reiners}}{{McLean}
  et~al.}{2012}]{McLean2012ApJ...746...23M}
{McLean} M.,  {Berger} E.,   {Reiners} A.,  2012, \mn@doi [\apj]
  {10.1088/0004-637X/746/1/23}, \href
  {http://adsabs.harvard.edu/abs/2012ApJ...746...23M} {746, 23}

\bibitem[\protect\citeauthoryear{{Medina}, {Winters}, {Irwin}  \&
  {Charbonneau}}{{Medina} et~al.}{2020}]{Medina2020ApJ...905..107M}
{Medina} A.~A.,  {Winters} J.~G.,  {Irwin} J.~M.,   {Charbonneau} D.,  2020,
  \mn@doi [\apj] {10.3847/1538-4357/abc686}, \href
  {https://ui.adsabs.harvard.edu/abs/2020ApJ...905..107M} {905, 107}

\bibitem[\protect\citeauthoryear{{Moe} \& {Kratter}}{{Moe} \&
  {Kratter}}{2019}]{Moe2019arXiv191201699M}
{Moe} M.,  {Kratter} K.~M.,  2019, arXiv e-prints, \href
  {https://ui.adsabs.harvard.edu/abs/2019arXiv191201699M} {p. arXiv:1912.01699}

\bibitem[\protect\citeauthoryear{{Morgan}, {West}, {Garc{\'e}s}, {Catal{\'a}n},
  {Dhital}, {Fuchs}  \& {Silvestri}}{{Morgan}
  et~al.}{2012}]{Morgan2012AJ....144...93M}
{Morgan} D.~P.,  {West} A.~A.,  {Garc{\'e}s} A.,  {Catal{\'a}n} S.,  {Dhital}
  S.,  {Fuchs} M.,   {Silvestri} N.~M.,  2012, \mn@doi [\aj]
  {10.1088/0004-6256/144/4/93}, \href
  {https://ui.adsabs.harvard.edu/abs/2012AJ....144...93M} {144, 93}

\bibitem[\protect\citeauthoryear{{Morgan}, {West}  \& {Becker}}{{Morgan}
  et~al.}{2016}]{Morgan2016AJ....151..114M}
{Morgan} D.~P.,  {West} A.~A.,   {Becker} A.~C.,  2016, \mn@doi [\aj]
  {10.3847/0004-6256/151/5/114}, \href
  {https://ui.adsabs.harvard.edu/abs/2016AJ....151..114M} {151, 114}

\bibitem[\protect\citeauthoryear{{Osten} \& {Jayawardhana}}{{Osten} \&
  {Jayawardhana}}{2006}]{OstenJayawardhana2006ApJ...644L..67O}
{Osten} R.~A.,  {Jayawardhana} R.,  2006, \mn@doi [\apjl] {10.1086/505328},
  \href {https://ui.adsabs.harvard.edu/abs/2006ApJ...644L..67O} {644, L67}

\bibitem[\protect\citeauthoryear{{Osten}, {Hawley}, {Bastian}  \&
  {Reid}}{{Osten} et~al.}{2006}]{Osten2006ApJ...637..518O}
{Osten} R.~A.,  {Hawley} S.~L.,  {Bastian} T.~S.,   {Reid} I.~N.,  2006,
  \mn@doi [\apj] {10.1086/498345}, \href
  {https://ui.adsabs.harvard.edu/abs/2006ApJ...637..518O} {637, 518}

\bibitem[\protect\citeauthoryear{{Osten}, {Melis}, {Stelzer}, {Bannister},
  {Radigan}, {Burgasser}, {Wolszczan}  \& {Luhman}}{{Osten}
  et~al.}{2015}]{Osten2015ApJ...805L...3O}
{Osten} R.~A.,  {Melis} C.,  {Stelzer} B.,  {Bannister} K.~W.,  {Radigan} J.,
  {Burgasser} A.~J.,  {Wolszczan} A.,   {Luhman} K.~L.,  2015, \mn@doi [\apjl]
  {10.1088/2041-8205/805/1/L3}, \href
  {https://ui.adsabs.harvard.edu/abs/2015ApJ...805L...3O} {805, L3}

\bibitem[\protect\citeauthoryear{{Owocki}, {Shultz}, {ud-Doula}, {Chandra},
  {Das}  \& {Leto}}{{Owocki} et~al.}{2022}]{Owocki2022MNRAS.513.1449O}
{Owocki} S.~P.,  {Shultz} M.~E.,  {ud-Doula} A.,  {Chandra} P.,  {Das} B.,
  {Leto} P.,  2022, \mn@doi [\mnras] {10.1093/mnras/stac341}, \href
  {https://ui.adsabs.harvard.edu/abs/2022MNRAS.513.1449O} {513, 1449}

\bibitem[\protect\citeauthoryear{{Paudel}, {Gizis}, {Mullan}, {Schmidt},
  {Burgasser}, {Williams}  \& {Berger}}{{Paudel}
  et~al.}{2018a}]{Paudel2018ApJ...858...55P}
{Paudel} R.~R.,  {Gizis} J.~E.,  {Mullan} D.~J.,  {Schmidt} S.~J.,  {Burgasser}
  A.~J.,  {Williams} P. K.~G.,   {Berger} E.,  2018a, \mn@doi [\apj]
  {10.3847/1538-4357/aab8fe}, \href
  {https://ui.adsabs.harvard.edu/abs/2018ApJ...858...55P} {858, 55}

\bibitem[\protect\citeauthoryear{{Paudel}, {Gizis}, {Mullan}, {Schmidt},
  {Burgasser}, {Williams}  \& {Berger}}{{Paudel}
  et~al.}{2018b}]{Paudel2018ApJ...861...76P}
{Paudel} R.~R.,  {Gizis} J.~E.,  {Mullan} D.~J.,  {Schmidt} S.~J.,  {Burgasser}
  A.~J.,  {Williams} P. K.~G.,   {Berger} E.,  2018b, \mn@doi [\apj]
  {10.3847/1538-4357/aac8e0}, \href
  {https://ui.adsabs.harvard.edu/abs/2018ApJ...861...76P} {861, 76}

\bibitem[\protect\citeauthoryear{{Paudel}, {Gizis}, {Mullan}, {Schmidt},
  {Burgasser}  \& {Williams}}{{Paudel}
  et~al.}{2020}]{Paudel2020MNRAS.494.5751P}
{Paudel} R.~R.,  {Gizis} J.~E.,  {Mullan} D.~J.,  {Schmidt} S.~J.,  {Burgasser}
  A.~J.,   {Williams} P.~K.~G.,  2020, \mn@doi [\mnras]
  {10.1093/mnras/staa1137}, \href
  {https://ui.adsabs.harvard.edu/abs/2020MNRAS.494.5751P} {494, 5751}

\bibitem[\protect\citeauthoryear{Pedregosa et~al.,}{Pedregosa
  et~al.}{2011}]{scikit-learn}
Pedregosa F.,  et~al., 2011, Journal of Machine Learning Research, 12, 2825

\bibitem[\protect\citeauthoryear{{Pineda} \& {Hallinan}}{{Pineda} \&
  {Hallinan}}{2018}]{Pineda2018ApJ...866..155P}
{Pineda} J.~S.,  {Hallinan} G.,  2018, \mn@doi [\apj]
  {10.3847/1538-4357/aae078}, \href
  {https://ui.adsabs.harvard.edu/abs/2018ApJ...866..155P} {866, 155}

\bibitem[\protect\citeauthoryear{{Pineda}, {Hallinan}  \& {Kao}}{{Pineda}
  et~al.}{2017}]{Pineda2017ApJ...846...75P}
{Pineda} J.~S.,  {Hallinan} G.,   {Kao} M.~M.,  2017, \mn@doi [\apj]
  {10.3847/1538-4357/aa8596}, \href
  {http://adsabs.harvard.edu/abs/2017ApJ...846...75P} {846, 75}

\bibitem[\protect\citeauthoryear{{Pineda}, {Youngblood}  \& {France}}{{Pineda}
  et~al.}{2021}]{pineda2021ApJ...918...40P}
{Pineda} J.~S.,  {Youngblood} A.,   {France} K.,  2021, \mn@doi [\apj]
  {10.3847/1538-4357/ac0aea}, \href
  {https://ui.adsabs.harvard.edu/abs/2021ApJ...918...40P} {918, 40}

\bibitem[\protect\citeauthoryear{{Potter}, {Mart{\'\i}n}, {Cushing}, {Baudoz},
  {Brandner}, {Guyon}  \& {Neuh{\"a}user}}{{Potter}
  et~al.}{2002}]{Potter2002ApJ...567L.133P}
{Potter} D.,  {Mart{\'\i}n} E.~L.,  {Cushing} M.~C.,  {Baudoz} P.,  {Brandner}
  W.,  {Guyon} O.,   {Neuh{\"a}user} R.,  2002, \mn@doi [\apjl]
  {10.1086/339999}, \href
  {https://ui.adsabs.harvard.edu/abs/2002ApJ...567L.133P} {567, L133}

\bibitem[\protect\citeauthoryear{{Rafikov} \& {Silsbee}}{{Rafikov} \&
  {Silsbee}}{2015}]{Rafikov2015ApJ...798...69R}
{Rafikov} R.~R.,  {Silsbee} K.,  2015, \mn@doi [\apj]
  {10.1088/0004-637X/798/2/69}, \href
  {https://ui.adsabs.harvard.edu/abs/2015ApJ...798...69R} {798, 69}

\bibitem[\protect\citeauthoryear{{Reid}, {Kirkpatrick}, {Liebert}, {Gizis},
  {Dahn}  \& {Monet}}{{Reid} et~al.}{2002}]{Reid2002AJ....124..519R}
{Reid} I.~N.,  {Kirkpatrick} J.~D.,  {Liebert} J.,  {Gizis} J.~E.,  {Dahn}
  C.~C.,   {Monet} D.~G.,  2002, \mn@doi [\aj] {10.1086/340805}, \href
  {https://ui.adsabs.harvard.edu/abs/2002AJ....124..519R} {124, 519}

\bibitem[\protect\citeauthoryear{{Reid}, {Cruz}, {Kirkpatrick}, {Allen},
  {Mungall}, {Liebert}, {Lowrance}  \& {Sweet}}{{Reid}
  et~al.}{2008}]{Reid2008AJ....136.1290R}
{Reid} I.~N.,  {Cruz} K.~L.,  {Kirkpatrick} J.~D.,  {Allen} P.~R.,  {Mungall}
  F.,  {Liebert} J.,  {Lowrance} P.,   {Sweet} A.,  2008, \mn@doi [\aj]
  {10.1088/0004-6256/136/3/1290}, \href
  {https://ui.adsabs.harvard.edu/abs/2008AJ....136.1290R} {136, 1290}

\bibitem[\protect\citeauthoryear{{Richey-Yowell}, {Kao}, {Pineda}, {Shkolnik}
  \& {Hallinan}}{{Richey-Yowell} et~al.}{2020}]{Richey-Yowell2020}
{Richey-Yowell} T.,  {Kao} M.~M.,  {Pineda} J.~S.,  {Shkolnik} E.~L.,
  {Hallinan} G.,  2020, \mn@doi [\apj] {10.3847/1538-4357/abb826}, \href
  {https://ui.adsabs.harvard.edu/abs/2020ApJ...903...74R} {903, 74}

\bibitem[\protect\citeauthoryear{{Rose} et~al.,}{{Rose}
  et~al.}{2023}]{Rose2023ApJ...951L..43R}
{Rose} K.,  et~al., 2023, \mn@doi [\apjl] {10.3847/2041-8213/ace188}, \href
  {https://ui.adsabs.harvard.edu/abs/2023ApJ...951L..43R} {951, L43}

\bibitem[\protect\citeauthoryear{{Route} \& {Wolszczan}}{{Route} \&
  {Wolszczan}}{2012}]{RouteWolszczan2012ApJ...747L..22R}
{Route} M.,  {Wolszczan} A.,  2012, \mn@doi [\apjl]
  {10.1088/2041-8205/747/2/L22}, \href
  {http://adsabs.harvard.edu/abs/2012ApJ...747L..22R} {747, L22}

\bibitem[\protect\citeauthoryear{{Route} \& {Wolszczan}}{{Route} \&
  {Wolszczan}}{2016a}]{RouteWolszczan2016ApJ...821L..21R}
{Route} M.,  {Wolszczan} A.,  2016a, \mn@doi [\apjl]
  {10.3847/2041-8205/821/2/L21}, \href
  {http://adsabs.harvard.edu/abs/2016ApJ...821L..21R} {821, L21}

\bibitem[\protect\citeauthoryear{{Route} \& {Wolszczan}}{{Route} \&
  {Wolszczan}}{2016b}]{RouteWolszczan2016ApJ...830...85R}
{Route} M.,  {Wolszczan} A.,  2016b, \mn@doi [\apj]
  {10.3847/0004-637X/830/2/85}, \href
  {http://adsabs.harvard.edu/abs/2016ApJ...830...85R} {830, 85}

\bibitem[\protect\citeauthoryear{{Sault}, {Oosterloo}, {Dulk}  \&
  {Leblanc}}{{Sault} et~al.}{1997}]{sault1997}
{Sault} R.~J.,  {Oosterloo} T.,  {Dulk} G.~A.,   {Leblanc} Y.,  1997, \aap,
  \href {http://adsabs.harvard.edu/abs/1997A%26A...324.1190S} {324, 1190}

\bibitem[\protect\citeauthoryear{{Schmidt}, {Cruz}, {Bongiorno}, {Liebert}  \&
  {Reid}}{{Schmidt} et~al.}{2007}]{Schmidt2007AJ....133.2258S}
{Schmidt} S.~J.,  {Cruz} K.~L.,  {Bongiorno} B.~J.,  {Liebert} J.,   {Reid}
  I.~N.,  2007, \mn@doi [\aj] {10.1086/512158}, \href
  {https://ui.adsabs.harvard.edu/abs/2007AJ....133.2258S} {133, 2258}

\bibitem[\protect\citeauthoryear{{Schmidt}, {Hawley}, {West}, {Bochanski},
  {Davenport}, {Ge}  \& {Schneider}}{{Schmidt}
  et~al.}{2015}]{Schmidt2015AJ....149..158S}
{Schmidt} S.~J.,  {Hawley} S.~L.,  {West} A.~A.,  {Bochanski} J.~J.,
  {Davenport} J. R.~A.,  {Ge} J.,   {Schneider} D.~P.,  2015, \mn@doi [\aj]
  {10.1088/0004-6256/149/5/158}, \href
  {https://ui.adsabs.harvard.edu/abs/2015AJ....149..158S} {149, 158}

\bibitem[\protect\citeauthoryear{{Scholz}, {McCaughrean}, {Lodieu}  \&
  {Kuhlbrodt}}{{Scholz} et~al.}{2003}]{Scholz2003AA...398L..29S}
{Scholz} R.~D.,  {McCaughrean} M.~J.,  {Lodieu} N.,   {Kuhlbrodt} B.,  2003,
  \mn@doi [\aap] {10.1051/0004-6361:20021847}, \href
  {https://ui.adsabs.harvard.edu/abs/2003A&A...398L..29S} {398, L29}

\bibitem[\protect\citeauthoryear{{Scholz}, {Moore}, {Jayawardhana}, {Aigrain},
  {Peterson}  \& {Stelzer}}{{Scholz} et~al.}{2018}]{Scholz2018ApJ...859..153S}
{Scholz} A.,  {Moore} K.,  {Jayawardhana} R.,  {Aigrain} S.,  {Peterson} D.,
  {Stelzer} B.,  2018, \mn@doi [\apj] {10.3847/1538-4357/aabfbe}, \href
  {https://ui.adsabs.harvard.edu/abs/2018ApJ...859..153S} {859, 153}

\bibitem[\protect\citeauthoryear{{Siegler}, {Close}, {Cruz}, {Mart{\'\i}n}  \&
  {Reid}}{{Siegler} et~al.}{2005}]{Siegler2005ApJ...621.1023S}
{Siegler} N.,  {Close} L.~M.,  {Cruz} K.~L.,  {Mart{\'\i}n} E.~L.,   {Reid}
  I.~N.,  2005, \mn@doi [\apj] {10.1086/427743}, \href
  {https://ui.adsabs.harvard.edu/abs/2005ApJ...621.1023S} {621, 1023}

\bibitem[\protect\citeauthoryear{{Smart} et~al.,}{{Smart}
  et~al.}{2013}]{Smart2013MNRAS.433.2054S}
{Smart} R.~L.,  et~al., 2013, \mn@doi [\mnras] {10.1093/mnras/stt876}, \href
  {https://ui.adsabs.harvard.edu/abs/2013MNRAS.433.2054S} {433, 2054}

\bibitem[\protect\citeauthoryear{{Sofue}, {Nakanishi}  \& {Ichiki}}{{Sofue}
  et~al.}{2019}]{Sofue2019MNRAS.485..924S}
{Sofue} Y.,  {Nakanishi} H.,   {Ichiki} K.,  2019, \mn@doi [\mnras]
  {10.1093/mnras/stz407}, \href
  {https://ui.adsabs.harvard.edu/abs/2019MNRAS.485..924S} {485, 924}

\bibitem[\protect\citeauthoryear{{Stone} et~al.,}{{Stone}
  et~al.}{2016}]{Stone2016ApJ...818L..12S}
{Stone} J.~M.,  et~al., 2016, \mn@doi [\apjl] {10.3847/2041-8205/818/1/L12},
  \href {https://ui.adsabs.harvard.edu/abs/2016ApJ...818L..12S} {818, L12}

\bibitem[\protect\citeauthoryear{{Stumpf}, {Brandner}, {Bouy}, {Henning}  \&
  {Hippler}}{{Stumpf} et~al.}{2010}]{Stumpf2010AA...516A..37S}
{Stumpf} M.~B.,  {Brandner} W.,  {Bouy} H.,  {Henning} T.,   {Hippler} S.,
  2010, \mn@doi [\aap] {10.1051/0004-6361/200913711}, \href
  {https://ui.adsabs.harvard.edu/abs/2010A&A...516A..37S} {516, A37}

\bibitem[\protect\citeauthoryear{{Tinney}, {Faherty}, {Kirkpatrick}, {Cushing},
  {Morley}  \& {Wright}}{{Tinney} et~al.}{2014}]{Tinney2014ApJ...796...39T}
{Tinney} C.~G.,  {Faherty} J.~K.,  {Kirkpatrick} J.~D.,  {Cushing} M.,
  {Morley} C.~V.,   {Wright} E.~L.,  2014, \mn@doi [\apj]
  {10.1088/0004-637X/796/1/39}, \href
  {https://ui.adsabs.harvard.edu/abs/2014ApJ...796...39T} {796, 39}

\bibitem[\protect\citeauthoryear{{Vedantham} et~al.,}{{Vedantham}
  et~al.}{2020}]{Vedantham2020ApJ...903L..33V}
{Vedantham} H.~K.,  et~al., 2020, \mn@doi [\apjl] {10.3847/2041-8213/abc256},
  \href {https://ui.adsabs.harvard.edu/abs/2020ApJ...903L..33V} {903, L33}

\bibitem[\protect\citeauthoryear{{Vedantham} et~al.,}{{Vedantham}
  et~al.}{2023}]{Vedantham2023A&A...675L...6V}
{Vedantham} H.~K.,  et~al., 2023, \mn@doi [\aap] {10.1051/0004-6361/202244965},
  \href {https://ui.adsabs.harvard.edu/abs/2023A&A...675L...6V} {675, L6}

\bibitem[\protect\citeauthoryear{{Villadsen} \& {Hallinan}}{{Villadsen} \&
  {Hallinan}}{2019}]{Villadsen2019ApJ...871..214V}
{Villadsen} J.,  {Hallinan} G.,  2019, \mn@doi [\apj]
  {10.3847/1538-4357/aaf88e}, \href
  {https://ui.adsabs.harvard.edu/abs/2019ApJ...871..214V} {871, 214}

\bibitem[\protect\citeauthoryear{{Villadsen}, {Hallinan}, {Bourke}, {G{\"u}del}
   \& {Rupen}}{{Villadsen} et~al.}{2014}]{Villadsen2014ApJ...788..112V}
{Villadsen} J.,  {Hallinan} G.,  {Bourke} S.,  {G{\"u}del} M.,   {Rupen} M.,
  2014, \mn@doi [\apj] {10.1088/0004-637X/788/2/112}, \href
  {https://ui.adsabs.harvard.edu/abs/2014ApJ...788..112V} {788, 112}

\bibitem[\protect\citeauthoryear{Virtanen et~al.,}{Virtanen
  et~al.}{2020}]{scipy}
Virtanen P.,  et~al., 2020, \mn@doi [Nature Methods]
  {10.1038/s41592-019-0686-2}, \href {https://rdcu.be/b08Wh} {17, 261}

\bibitem[\protect\citeauthoryear{{Weinberger}, {Boss}, {Keiser},
  {Anglada-Escud{\'e}}, {Thompson}  \& {Burley}}{{Weinberger}
  et~al.}{2016}]{Weinberger2016AJ....152...24W}
{Weinberger} A.~J.,  {Boss} A.~P.,  {Keiser} S.~A.,  {Anglada-Escud{\'e}} G.,
  {Thompson} I.~B.,   {Burley} G.,  2016, \mn@doi [\aj]
  {10.3847/0004-6256/152/1/24}, \href
  {https://ui.adsabs.harvard.edu/abs/2016AJ....152...24W} {152, 24}

\bibitem[\protect\citeauthoryear{{Williams} \& {Berger}}{{Williams} \&
  {Berger}}{2015}]{Williams2015ApJ...808..189W}
{Williams} P.~K.~G.,  {Berger} E.,  2015, \mn@doi [\apj]
  {10.1088/0004-637X/808/2/189}, \href
  {http://adsabs.harvard.edu/abs/2015ApJ...808..189W} {808, 189}

\bibitem[\protect\citeauthoryear{{Williams}, {Cook}  \& {Berger}}{{Williams}
  et~al.}{2014}]{Williams2014ApJ...785....9W}
{Williams} P.~K.~G.,  {Cook} B.~A.,   {Berger} E.,  2014, \mn@doi [\apj]
  {10.1088/0004-637X/785/1/9}, \href
  {http://adsabs.harvard.edu/abs/2014ApJ...785....9W} {785, 9}

\bibitem[\protect\citeauthoryear{{Williams}, {Berger}, {Irwin},
  {Berta-Thompson}  \& {Charbonneau}}{{Williams}
  et~al.}{2015a}]{Williams2015ApJ...799..192W}
{Williams} P.~K.~G.,  {Berger} E.,  {Irwin} J.,  {Berta-Thompson} Z.~K.,
  {Charbonneau} D.,  2015a, \mn@doi [\apj] {10.1088/0004-637X/799/2/192}, \href
  {https://ui.adsabs.harvard.edu/abs/2015ApJ...799..192W} {799, 192}

\bibitem[\protect\citeauthoryear{{Williams}, {Casewell}, {Stark}, {Littlefair},
  {Helling}  \& {Berger}}{{Williams}
  et~al.}{2015b}]{Williams2015ApJ...815...64W}
{Williams} P.~K.~G.,  {Casewell} S.~L.,  {Stark} C.~R.,  {Littlefair} S.~P.,
  {Helling} C.,   {Berger} E.,  2015b, \mn@doi [\apj]
  {10.1088/0004-637X/815/1/64}, \href
  {http://adsabs.harvard.edu/abs/2015ApJ...815...64W} {815, 64}

\bibitem[\protect\citeauthoryear{{Williams}, {Gizis}  \& {Berger}}{{Williams}
  et~al.}{2017}]{Williams2017ApJ...834..117W}
{Williams} P.~K.~G.,  {Gizis} J.~E.,   {Berger} E.,  2017, \mn@doi [\apj]
  {10.3847/1538-4357/834/2/117}, \href
  {http://adsabs.harvard.edu/abs/2017ApJ...834..117W} {834, 117}

\bibitem[\protect\citeauthoryear{{Zahn}}{{Zahn}}{1977}]{Zahn1977A&A....57..383Z}
{Zahn} J.~P.,  1977, \aap, \href
  {https://ui.adsabs.harvard.edu/abs/1977A&A....57..383Z} {500, 121}

\bibitem[\protect\citeauthoryear{{Zhang}, {Hallinan}, {Brisken}, {Bourke}  \&
  {Golden}}{{Zhang} et~al.}{2020}]{Zhang2020ApJ...897...11Z}
{Zhang} Q.,  {Hallinan} G.,  {Brisken} W.,  {Bourke} S.,   {Golden} A.,  2020,
  \mn@doi [\apj] {10.3847/1538-4357/ab9177}, \href
  {https://ui.adsabs.harvard.edu/abs/2020ApJ...897...11Z} {897, 11}

\bibitem[\protect\citeauthoryear{{van Haarlem} et~al.,}{{van Haarlem}
  et~al.}{2013}]{vanHaarlem2013AA...556A...2V}
{van Haarlem} M.~P.,  et~al., 2013, \mn@doi [\aap]
  {10.1051/0004-6361/201220873}, \href
  {https://ui.adsabs.harvard.edu/abs/2013A&A...556A...2V} {556, A2}

\bibitem[\protect\citeauthoryear{{van der Walt}, {Colbert}  \&
  {Varoquaux}}{{van der Walt} et~al.}{2011}]{numpy2}
{van der Walt} S.,  {Colbert} S.~C.,   {Varoquaux} G.,  2011, \mn@doi
  [Computing in Science and Engineering] {10.1109/MCSE.2011.37}, \href
  {http://adsabs.harvard.edu/abs/2011CSE....13b..22V} {13, 22}

\makeatother
\end{thebibliography}


\appendix
\section{Proof for Equation 9}     \label{sec:proof}

\begin{align} 
   \theta_{\predict} 	& = \int_{0}^{L_{\mathrm{max}}} \prob(L_a  \geq L_{\mathrm{min}} - k) \PDF(L_b=k) \, dk \quad  \\
   				& = \int_{0}^{L_{\mathrm{min}}} \prob(L_a  \geq L_{\mathrm{min}} - k) \PDF(L_b=k) \, dk \quad + \quad  \int_{L_{\mathrm{min}}}^{L_{\mathrm{max}}} \prob(L_a  \geq L_{\mathrm{min}} - k) \PDF(L_b=k) \, dk \quad.
\end{align}

Substitution by equation \ref{eqn:assumption}  together with the insight that the combined light of a binary system must be more luminous than $L_{\mathrm{min}}$ when $k > L_{\mathrm{min}}$ gives
\begin{align} \label{eqn:integrand}
   \theta_{\predict} 	& = (1-\theta_{\mathrm{single}}) \int_{0}^{L_{\mathrm{min}}} \prob(L_a  \geq L_{\mathrm{min}} - k)  \mathcal{F}_2(L_{b} = k)   \, dk \quad + \quad  \theta_{\mathrm{single}} \quad.
\end{align}
This integrand is a joint probability distribution over the random variables $L_a$ and  $L_b$ where the latter constrains the set of possible values for the former. This allows equation \ref{eqn:integrand} to be rewritten as 
\begin{align} 
   \theta_{\predict}    & = \theta_{\mathrm{single}} + (1-\theta_{\mathrm{single}}) \int_{k=0}^{L_{\mathrm{min}}}    \int_{x= L_{\mathrm{min}} - k}^{\infty}  \PDF(L_a = x )  \mathcal{F}_2(L_b = k)   \, dx  \, dk \\ 
                & = \theta_{\mathrm{single}} + (1-\theta_{\mathrm{single}}) \int_{k=0}^{L_{\mathrm{min}}}    \mathcal{F}_2(L_b = k)  \,     \int_{x= L_{\mathrm{min}} - k}^{\infty}  \PDF(L_a = x ) \, dx \, dk \\
				& = \theta_{\mathrm{single}} + (1-\theta_{\mathrm{single}}) \int_{k=0}^{L_{\mathrm{min}}}    \mathcal{F}_2(L_b = k) 
					 \left[   	\int_{x= L_{\mathrm{min}} - k}^{L_{\mathrm{min}}} \PDF(L_a = x ) \, dx 	\, +  \,
					 		\int_{x= L_{\mathrm{min}}}^{L_{\mathrm{max}}} \PDF(L_a = x ) \, dx 		\, +  \,
							\int_{x= L_{\mathrm{max}}}^{\infty}  \PDF(L_a = x ) \, dx \right]\, dk \\
				& = \theta_{\mathrm{single}} + (1-\theta_{\mathrm{single}}) \int_{k=0}^{L_{\mathrm{min}}}    \mathcal{F}_2(L_b = k) 
					 \left[   	\int_{x= L_{\mathrm{min}} - k}^{L_{\mathrm{min}}} \PDF(L_a = x ) \, dx 	\quad +   \quad  \theta_{\mathrm{single}} \right]\, dk \\
				& = \theta_{\mathrm{single}} +(1-\theta_{\mathrm{single}})    
					 \left[ \int_{k=0}^{L_{\mathrm{min}}}    \mathcal{F}_2(L_b = k)   \int_{x= L_{\mathrm{min}} - k}^{L_{\mathrm{min}}} \PDF(L_a = x ) \, dx \, dk  \quad + \quad
					 \theta_{\mathrm{single}}   \right]
\end{align}
by substitution from equation  \ref{eqn:assumption}.  The first bracketed term evaluates to a constant $\mathbb{C}$ such that 
\begin{align}
	 \theta_{\text{\predict}} = \mathcal{C}(\theta_{\si}) + \big[  1- (1-\theta_{\si})^2 \big]  \quad,
\end{align}
where  $\mathcal{C}(\theta_{\si}) = \mathbb{C}(1-\theta_{\si})$ is function that depends on $\theta_{\si}$, as desired.

%


\bsp	
\label{lastpage}
\end{document}